\documentclass[%
  reprint,
  superscriptaddress,
 amsmath,amssymb,
 aps,
 prd,
 floatfix
]{revtex4-1}

\usepackage{graphicx}  
\usepackage{epstopdf}
\usepackage{dcolumn}   
\usepackage{bm}        
\usepackage{color}
\usepackage[caption=false]{subfig}
\usepackage{epsfig}
\usepackage{multirow}
\usepackage[]{units}
\usepackage{amssymb}   
\usepackage[capitalize]{cleveref}
\usepackage{hyperref}
\usepackage{placeins}

\RequirePackage{xspace}

\def\dcp      {\ensuremath{\delta_{\mathrm{CP}}}\xspace}

\def\nue        {\ensuremath{\nu_e}\xspace}

\def\num        {\ensuremath{\nu_\mu}\xspace}

\def\Kbar  {\kern 0.2em\overline{\kern -0.2em K}{}\xspace}

\def\Kz    {\ensuremath{K^0}\xspace}
\def\Kzb   {\ensuremath{\Kbar^0}\xspace}
\def\KzKzb {\ensuremath{\Kz \kern -0.16em \Kzb}\xspace}
\def\Kp    {\ensuremath{K^+}\xspace}
\def\Km    {\ensuremath{K^-}\xspace}

\def\KpKm  {\ensuremath{\Kp \kern -0.16em \Km}\xspace}

\newcommand{\tev}{\ensuremath{\mathrm{\,Te\kern -0.1em V}}\xspace}
\newcommand{\gev}{\ensuremath{\mathrm{\,Ge\kern -0.1em V}}\xspace}
\newcommand{\mev}{\ensuremath{\mathrm{\,Me\kern -0.1em V}}\xspace}
\newcommand{\kev}{\ensuremath{\mathrm{\,ke\kern -0.1em V}}\xspace}
\newcommand{\ev}{\ensuremath{\mathrm{\,e\kern -0.1em V}}\xspace}
\newcommand{\gevc}{\ensuremath{{\mathrm{\,Ge\kern -0.1em V\!/}c}}\xspace}
\newcommand{\mevc}{\ensuremath{{\mathrm{\,Me\kern -0.1em V\!/}c}}\xspace}
\newcommand{\gevcc}{\ensuremath{{\mathrm{\,Ge\kern -0.1em V\!/}c^2}}\xspace}
\newcommand{\mevcc}{\ensuremath{{\mathrm{\,Me\kern -0.1em V\!/}c^2}}\xspace}

\def\mus  {\ensuremath{\rm \,\mus}\xspace}

\def\mus        {\ensuremath{\,\mu{\rm s}}\xspace}    

\newcommand{\podd}{P\O{}D\xspace}

\newcommand{\pizero}{$\pi^0$\xspace}

\newcommand{\ccnue}{CC$\nu_{e}$\xspace}

\begin{document}

\preprint{XXX}

\title{Measurement of the Electron Neutrino Charged-current
  Interaction Rate on Water with the T2K ND280 $\pi^0$ Detector}

\newcommand{\kkoc}{\affiliation{University of Alberta, Centre for Particle Physics, Department of Physics, Edmonton, Alberta, Canada}}
\newcommand{\ocaf}{\affiliation{BMCC/CUNY, Science Department, New York, New York, U.S.A.}}
\newcommand{\koee}{\affiliation{University of Bern, Albert Einstein Center for Fundamental Physics, Laboratory for High Energy Physics (LHEP), Bern, Switzerland}}
\newcommand{\kofe}{\affiliation{Boston University, Department of Physics, Boston, Massachusetts, U.S.A.}}
\newcommand{\kkod}{\affiliation{University of British Columbia, Department of Physics and Astronomy, Vancouver, British Columbia, Canada}}
\newcommand{\koga}{\affiliation{University of California, Irvine, Department of Physics and Astronomy, Irvine, California, U.S.A.}}
\newcommand{\kofg}{\affiliation{Colorado State University, Department of Physics, Fort Collins, Colorado, U.S.A.}}
\newcommand{\kogb}{\affiliation{University of Colorado at Boulder, Department of Physics, Boulder, Colorado, U.S.A.}}
\newcommand{\oehb}{\affiliation{Daresbury Laboratory, Warrington, United Kingdom}}
\newcommand{\kofh}{\affiliation{Duke University, Department of Physics, Durham, North Carolina, U.S.A.}}
\newcommand{\koef}{\affiliation{ETH Zurich, Institute for Particle Physics, Zurich, Switzerland}}
\newcommand{\koba}{\affiliation{Ecole Polytechnique, IN2P3-CNRS, Laboratoire Leprince-Ringuet, Palaiseau, France }}
\newcommand{\koeg}{\affiliation{University of Geneva, Section de Physique, DPNC, Geneva, Switzerland}}
\newcommand{\kodg}{\affiliation{H. Niewodniczanski Institute of Nuclear Physics PAN, Cracow, Poland}}
\newcommand{\kocb}{\affiliation{High Energy Accelerator Research Organization (KEK), Tsukuba, Ibaraki, Japan}}
\newcommand{\koec}{\affiliation{IFIC (CSIC \& University of Valencia), Valencia, Spain}}
\newcommand{\kogf}{\affiliation{INFN Sezione di Bari and Universit\`a e Politecnico di Bari, Dipartimento Interuniversitario di Fisica, Bari, Italy}}
\newcommand{\ogfa}{\affiliation{INFN Sezione di Bari, Bari, Italy}}
\newcommand{\kobe}{\affiliation{INFN Sezione di Napoli and Universit\`a di Napoli, Dipartimento di Fisica, Napoli, Italy}}
\newcommand{\obea}{\affiliation{INFN Sezione di Napoli, Napoli, Italy}}
\newcommand{\kobf}{\affiliation{INFN Sezione di Padova and Universit\`a di Padova, Dipartimento di Fisica, Padova, Italy}}
\newcommand{\obfa}{\affiliation{INFN Sezione di Padova, Padova, Italy}}
\newcommand{\kobd}{\affiliation{INFN Sezione di Roma and Universit\`a di Roma ``La Sapienza'', Roma, Italy}}
\newcommand{\obda}{\affiliation{INFN Sezione di Roma, Roma, Italy}}
\newcommand{\kkoi}{\affiliation{IRFU, CEA Saclay, Gif-sur-Yvette, France}}
\newcommand{\koei}{\affiliation{Imperial College London, Department of Physics, London, United Kingdom}}
\newcommand{\koed}{\affiliation{Institut de Fisica d'Altes Energies (IFAE), Bellaterra (Barcelona), Spain}}
\newcommand{\koeb}{\affiliation{Institute for Nuclear Research of the Russian Academy of Sciences, Moscow, Russia}}
\newcommand{\ocae}{\affiliation{Institute of Particle Physics, Canada}}
\newcommand{\ocab}{\affiliation{J-PARC, Tokai, Japan}}
\newcommand{\ocad}{\affiliation{JINR, Dubna, Russia}}
\newcommand{\ocag}{\affiliation{Kavli IPMU (WPI), the University of Tokyo, Japan}}
\newcommand{\koha}{\affiliation{Kavli Institute for the Physics and Mathematics of the Universe (WPI), Todai Institutes for Advanced Study, University of Tokyo, Kashiwa, Chiba, Japan}}
\newcommand{\kocc}{\affiliation{Kobe University, Kobe, Japan}}
\newcommand{\kocd}{\affiliation{Kyoto University, Department of Physics, Kyoto, Japan}}
\newcommand{\koej}{\affiliation{Lancaster University, Physics Department, Lancaster, United Kingdom}}
\newcommand{\kofc}{\affiliation{University of Liverpool, Department of Physics, Liverpool, United Kingdom}}
\newcommand{\kofi}{\affiliation{Louisiana State University, Department of Physics and Astronomy, Baton Rouge, Louisiana, U.S.A.}}
\newcommand{\kohb}{\affiliation{Michigan State University, Department of Physics and Astronomy,  East Lansing, Michigan, U.S.A.}}
\newcommand{\koce}{\affiliation{Miyagi University of Education, Department of Physics, Sendai, Japan}}
\newcommand{\ocah}{\affiliation{Moscow Institute of Physics and Technology and National Research Nuclear University "MEPhI", Moscow, Russia}}
\newcommand{\kodf}{\affiliation{National Centre for Nuclear Research, Warsaw, Poland}}
\newcommand{\kogj}{\affiliation{Okayama University, Department of Physics, Okayama, Japan}}
\newcommand{\kocf}{\affiliation{Osaka City University, Department of Physics, Osaka, Japan}}
\newcommand{\kogg}{\affiliation{Oxford University, Department of Physics, Oxford, United Kingdom}}
\newcommand{\kogc}{\affiliation{University of Pittsburgh, Department of Physics and Astronomy, Pittsburgh, Pennsylvania, U.S.A.}}
\newcommand{\kofa}{\affiliation{Queen Mary University of London, School of Physics and Astronomy, London, United Kingdom}}
\newcommand{\kobc}{\affiliation{RWTH Aachen University, III. Physikalisches Institut, Aachen, Germany}}
\newcommand{\kkoe}{\affiliation{University of Regina, Department of Physics, Regina, Saskatchewan, Canada}}
\newcommand{\kogd}{\affiliation{University of Rochester, Department of Physics and Astronomy, Rochester, New York, U.S.A.}}
\newcommand{\koeh}{\affiliation{STFC, Rutherford Appleton Laboratory, Harwell Oxford,  and  Daresbury Laboratory, Warrington, United Kingdom}}
\newcommand{\oeha}{\affiliation{STFC, Rutherford Appleton Laboratory, Harwell Oxford, United Kingdom}}
\newcommand{\kofb}{\affiliation{University of Sheffield, Department of Physics and Astronomy, Sheffield, United Kingdom}}
\newcommand{\kodi}{\affiliation{University of Silesia, Institute of Physics, Katowice, Poland}}
\newcommand{\kofj}{\affiliation{State University of New York at Stony Brook, Department of Physics and Astronomy, Stony Brook, New York, U.S.A.}}
\newcommand{\kkob}{\affiliation{TRIUMF, Vancouver, British Columbia, Canada}}
\newcommand{\kogi}{\affiliation{Tokyo Metropolitan University, Department of Physics, Tokyo, Japan}}
\newcommand{\koch}{\affiliation{University of Tokyo, Department of Physics, Tokyo, Japan}}
\newcommand{\kobj}{\affiliation{University of Tokyo, Institute for Cosmic Ray Research, Kamioka Observatory, Kamioka, Japan}}
\newcommand{\kocg}{\affiliation{University of Tokyo, Institute for Cosmic Ray Research, Research Center for Cosmic Neutrinos, Kashiwa, Japan}}
\newcommand{\kkof}{\affiliation{University of Toronto, Department of Physics, Toronto, Ontario, Canada}}
\newcommand{\kobb}{\affiliation{UPMC, Universit\'e Paris Diderot, CNRS/IN2P3, Laboratoire de Physique Nucl\'eaire et de Hautes Energies (LPNHE), Paris, France}}
\newcommand{\kkoj}{\affiliation{Universit\'e de Lyon, Universit\'e Claude Bernard Lyon 1, IPN Lyon (IN2P3), Villeurbanne, France}}
\newcommand{\kkog}{\affiliation{University of Victoria, Department of Physics and Astronomy, Victoria, British Columbia, Canada}}
\newcommand{\kodh}{\affiliation{Warsaw University of Technology, Institute of Radioelectronics, Warsaw, Poland}}
\newcommand{\kodj}{\affiliation{University of Warsaw, Faculty of Physics, Warsaw, Poland}}
\newcommand{\kofd}{\affiliation{University of Warwick, Department of Physics, Coventry, United Kingdom}}
\newcommand{\koge}{\affiliation{University of Washington, Department of Physics, Seattle, Washington, U.S.A.}}
\newcommand{\kogh}{\affiliation{University of Winnipeg, Department of Physics, Winnipeg, Manitoba, Canada}}
\newcommand{\koea}{\affiliation{Wroclaw University, Faculty of Physics and Astronomy, Wroclaw, Poland}}
\newcommand{\kkoh}{\affiliation{York University, Department of Physics and Astronomy, Toronto, Ontario, Canada}}
\newcommand{\obeb}{\affiliation{Universit\`a di Napoli, Dipartimento di Fisica, Napoli, Italy}}
\newcommand{\obfb}{\affiliation{Universit\`a di Padova, Dipartimento di Fisica, Padova, Italy}}
\newcommand{\obdb}{\affiliation{Universit\`a di Roma ``La Sapienza, Roma, Italy}}
\newcommand{\ogfb}{\affiliation{Universit\`a e Politecnico di Bari, Dipartimento Interuniversitario di Fisica, Bari, Italy}}
\kkoc
\ocaf
\koee
\kofe
\kkod
\koga
\kofg
\kogb
\oehb
\kofh
\koef
\koba
\koeg
\kodg
\kocb
\koec
\kogf
\ogfa
\kobe
\obea
\kobf
\obfa
\kobd
\obda
\kkoi
\koei
\koed
\koeb
\ocae
\ocab
\ocad
\ocag
\koha
\kocc
\kocd
\koej
\kofc
\kofi
\kohb
\koce
\ocah
\kodf
\kogj
\kocf
\kogg
\kogc
\kofa
\kobc
\kkoe
\kogd
\koeh
\oeha
\kofb
\kodi
\kofj
\kkob
\kogi
\koch
\kobj
\kocg
\kkof
\kobb
\kkoj
\kkog
\kodh
\kodj
\kofd
\koge
\kogh
\koea
\kkoh
\obeb
\obfb
\obdb
\ogfb
\author{K. Abe}\kobj
\author{J. Adam}\kofj
\author{H. Aihara}\koch\koha
\author{C. Andreopoulos}\koeh\kofc
\author{S. Aoki}\kocc
\author{A. Ariga}\koee
\author{S. Assylbekov}\kofg
\author{D. Autiero}\kkoj
\author{M. Barbi}\kkoe
\author{G.J. Barker}\kofd
% ...

\date{\today}

\begin{abstract}


This paper presents a measurement of the charged current interaction
rate of the electron neutrino beam component of the beam above
$1.5$~GeV using the large fiducial mass of the T2K $\pi^0$ detector.
The predominant portion of the $\nu_e$ flux ($\sim$85\%) at these energies
comes from kaon decays.  The measured ratio of the observed beam
interaction rate to the predicted rate in the detector with water
targets filled is 0.89 $\pm$ 0.08 (stat.)  $\pm$ 0.11 (sys.), and with
the water targets emptied is 0.90 $\pm$ 0.09 (stat.)  $\pm$ 0.13
(sys.). The ratio obtained for the interactions on water only from an
event subtraction method is 0.87 $\pm$ 0.33 (stat.)  $\pm$ 0.21
(sys.). This is the first measurement of the interaction rate of
electron neutrinos on water, which is particularly of interest to
experiments with water Cherenkov detectors.

\end{abstract}

\pacs{Valid PACS appear here}
\maketitle


\section{\label{sec:intro} Introduction}

This paper reports a measurement of the ratio of the charged current
$\nu_e$ event rate relative to the simulation with NEUT~\cite{NEUT}
event generator, version 4.1.4.2, for neutrino energies above 1.5~GeV
in the T2K beam.  The interaction rate of electron neutrinos on water
has never been measured at the neutrinos energies above 1.5~GeV, or at
any other energies.  The mean reconstructed energy of the selected
neutrinos in the analysis presented in this paper is 2.7~GeV.  The \nue
cross section has been measured on a liquid freon target for energies
between 1.5~GeV and 8~GeV by Gargamelle \cite{Blietschau:1978mu} and
on $^{12}$C for energies around 32~MeV at LANSCE
\cite{Auerbach:2001wg}.  Also at lower energies, the anti-electron
neutrino interactions have been measured by experiments near nuclear
reactors.  A review of neutrino cross section measurements can be
found in \cite{RevModPhys.84.1307}.

The T2K experiment~\cite{Abe:2011ks} was built with the primary goals
of precisely determining the oscillation parameter $\theta_{13}$ via
electron neutrino appearance, and of the parameters $\theta_{23}$ and
$\Delta m^2_{32}$ via muon neutrino disappearance. The predominantly
$\nu_{\mu}$ beam for these measurements is produced at the Japan
Proton Accelerator Research Complex (J-PARC) in Tokai. The neutrinos
from this beam are observed at a near detector, ND280, 
which is located 280~m downstream from the production target, where
the neutrinos are not expected to have been affected by
oscillations. The T2K far detector, Super-Kamiokande (SK), then
measures the muon and electron neutrinos (and anti-neutrinos) after
they have undergone a near maximal oscillation.

The oscillation probability for $\nu_\mu \rightarrow \nu_e$ depends on
the mixing parameter, $\theta_{13}$, and on sub-leading effects that
depend on the CP-violating phase, \dcp, and on the mass
hierarchy~\cite{PhysRevD.56.3093}.  T2K has already observed the
appearance of 28 \nue candidate events at the far detector with a
7.3~$\sigma$ significance over a background expectation of
$4.92\pm0.55$ events for $\theta_{13}=0$~\cite{Abe:2013hdq}. The
largest irreducible background for the appearance measurement comes
from the predicted 3.2 intrinsic $\nu_e$ beam events.

In T2K the \nue are expected to represent about 1.2\% of the total
neutrino flux~\cite{PhysRevD.87.012001}. The T2K \num beam is produced
by magnetic focusing of pions and kaons produced by the interaction of
a proton beam with a graphite target. The unavoidable \nue component
comes from the decay of muons from pion decay, and from kaon decay. In
any long-baseline neutrino experiment proposed to measure CP violation
and precisely measure neutrino oscillation parameters, the \nue
component of the beam will be the main source of
background~\cite{Abe:2011ts, Adams:2013qkq, Stahl:2012exa}.

The measurement of the beam \nue charged current (\ccnue) interactions
on a plastic scintillator and water target using ND280 tracker, was reported in\cite{T2Kbeamnue1}.
This paper reports a direct measurement of this component of the
charged current (CC) neutrino interactions in the ND280 $\pi^{0}$
Detector (\podd{})~\cite{Assylbekov201248}, which is located just
upstream of the tracker.  In this selection, the majority of the
electron neutrinos were produced in kaon decay, and have energies
above 1.5~GeV. The \podd{} detector has water targets that can be
filled or emptied.  Data were taken both with the targets filled to
create a water target (water configuration), and empty to leave just
air in place of the water target (air configuration).  With data in the
two configurations a subtraction analysis obtained the interaction
rate just on water.

Similar to the subtraction analysis presented here, a ratio analysis
has been conducted by the Minerva collaboration for 2-20~GeV
$\nu_{\mu}$ on C, Fe, and Pb compared to CH\cite{BGTice:2014}. A
subtraction analysis of the Minerva data is presented the thesis of
B.G. Tice \cite{BGTiceThesis}.  Apart from the Minerva measurements,
this appears to be the only other use of the subtraction analysis to
date in neutrino scattering experiments.

The $\nu_e$ and $\nu_{\mu}$ come from the same pion to muon to
electron decay chain, and lepton universality allows the expected rate
of $\nu_e$ to be constrained by measuring the much larger flux of
$\nu_{\mu}$.  Details concerning the T2K beam flux measurement, and
further information on recent measurements of $\nu_{\mu}$
interactions in the near detectors, can be found in
Ref. ~\cite{Day:2012gb}.

One of the systematic uncertainties in long baseline neutrino
oscillation measurements using water Cherenkov detectors comes from
model uncertainties in the meson exchange current for C versus for O.
Having measurements of neutrino interaction rates on water is
therefore important.  For a recent review of $\nu_{\mu}$ cross section
measurements on various nuclear targets refer to the
PDG\cite{Zeller:2014}.  The only measurements of $\nu_{\mu}$ neutrino
interactions on water were reported by the K2K experiment for
quasi-elastic interactions \cite{K2KQE}, and for reactions resulting
in pions in the final state \cite{K2KPi1,K2KPi2,K2KPi3,K2KPi4}.


The paper is organized as follows. In Section \ref{sec:Detector} the
\podd{} detector, used to do the measurement is described. The
electron selection, and expected backgrounds are then described in
Section \ref{sec:EventSelection}. The particle identification (PID) to
select electrons from muons in the \podd{} is a key component of this
measurement, and will be described further in the section on event
selection. The water subtraction method is then described in
\ref{sec:WaterSubtraction}. The detector, reconstruction, flux and
cross section systematic uncertainties in the measurement are reviewed
in Section \ref{sec:Systematics}. Finally the results of the rate
measurement are presented in Section \ref{sec:Results} and a summary
is in Section \ref{sec:Conclusion}.


\section{\label{sec:Detector}ND280 $\pi^0$ Detector}

The T2K ND280 $\pi^0$ Detector (\podd) is a scintillator based
tracking calorimeter optimized to measure neutral current $\pi^0$ in
the momentum range that contributes to backgrounds for $\nu_e$
appearance~\cite{Assylbekov201248}.  Refer to Fig.~4 of
~\cite{Assylbekov201248} for a picture of the \podd detector.  The
\podd is composed of layers of plastic scintillator alternating with
water targets and brass sheets or lead sheets. The \podd sits in front
of a tracking detector made up of two fine grain scintillator modules
which serve as active targets placed between three time projection
chambers. Both the \podd and tracking detector are surrounded by
electromagnetic calorimeters and are in a 0.2~T magnetic field.

The \podd is constructed using 40 scintillator modules, each module is
constructed with two perpendicular arrays of triangular scintillating
bars and is approximately 38~mm thick. The scintillator modules are
arranged in three regions. The most upstream and downstream regions of
the detector are composed of seven modules interleaved with 4.5~mm
thick sheets of stainless steel-clad lead that function as 4.9
radiation length electromagnetic calorimeters to improve the
containment of photons and electrons. The central region serves as a
target containing water. It has 25 water target layers that are 28~mm
thick sandwiched between 26 scintillator modules and 1.3~mm brass
sheets, positioned in between water targets and scintillator
layers. The target region has a fiducial mass of approximately 1900~kg
of water and 3570~kg of other materials.

The energy resolution of the \podd can be estimated from Monte Carlo
studies by calculating the difference between true and reconstructed
energy for many events. The energy resolution for electrons after the
selections described in \ref{sec:EventSelection}, is 16\%.

\section{\label{sec:EventSelection}Event Selection}

\subsection{\label{sec:Overview}Overview}

In this analysis, all the data collected between January 2010 and May
2013 except for very small fraction of Run III data, due to the
magnetic horn current decrease which caused a failure in good spill
pre-selection, are used.  The data are subdivided into different run
periods and \podd configurations as shown in Table~\ref{tab:DataPOT}.
The simulated data used in this analysis corresponds to ten times the
Protons on Target (POT) of the data, and reproduces the various
experimental conditions of the different data-taking periods.

\begin{table}
	\caption{Summary of T2K runs and the number of protons on
          target (POT) used in the analysis.}
	\label{tab:DataPOT}
	\begin{ruledtabular}
		\begin{tabular}{lcccc}
			T2K run & \podd Config. & Beam Power (kW)& POT ($\times 10^{19}$) \\
			\hline
			Run I   & Water & 50  & 2.96 \\
			Run II  & Water & 120 & 6.96 \\
			Run II  & Air   & 120 & 3.59 \\
			Run III & Air   & 178 & 13.5 \\
			Run IV  & Water & 178 & 16.5 \\
			Run IV  & Air   & 178 & 17.8 \\
			\hline
			Total   & Water &     & 26.4 \\
			        & Air   &     & 34.9 \\
		\end{tabular}
	\end{ruledtabular}
\end{table}

Neutrino interactions in ND280 are simulated with the NEUT~\cite{NEUT}
event generator, version 5.1.4.2.  The generator covers a range of
neutrino energy from several tens of MeV to hundreds of TeV and
simulates all the nuclear targets present in ND280.  In the simulated
data, neutrino interactions are generated outside and within the full
ND280 volume including all active and inactive material, providing
information to understand the signal and backgrounds from interactions
outside the ND280 fiducial volume.  The details of the simulation
process are described in~\cite{PhysRevD.89.092003}.

Simulation of products of the neutrino interactions in the \podd is
done using a GEANT 4.9.4 simulation \cite{GEANT4_1, GEANT4_2,
  GEANT4_3, GEANT4_4}.  The standard GEANT physics list for
electromagnetic interactions is used in the simulation.

The analysis uses two reconstructed objects, a track and a shower.
Within the \podd{} reconstruction algorithm, hits in \podd
scintillator layer associated with a reconstructed track classified as
an electromagnetic track (typically electrons or photons) are
forwarded to the shower reconstruction stage.  Hits associated with a
track that are classified as a light track (typically muon) or a heavy
track (typically proton) are not forwarded to the shower
reconstruction stage and cannot be reconstructed as a shower.

\begin{figure}
	\includegraphics[width=\linewidth]{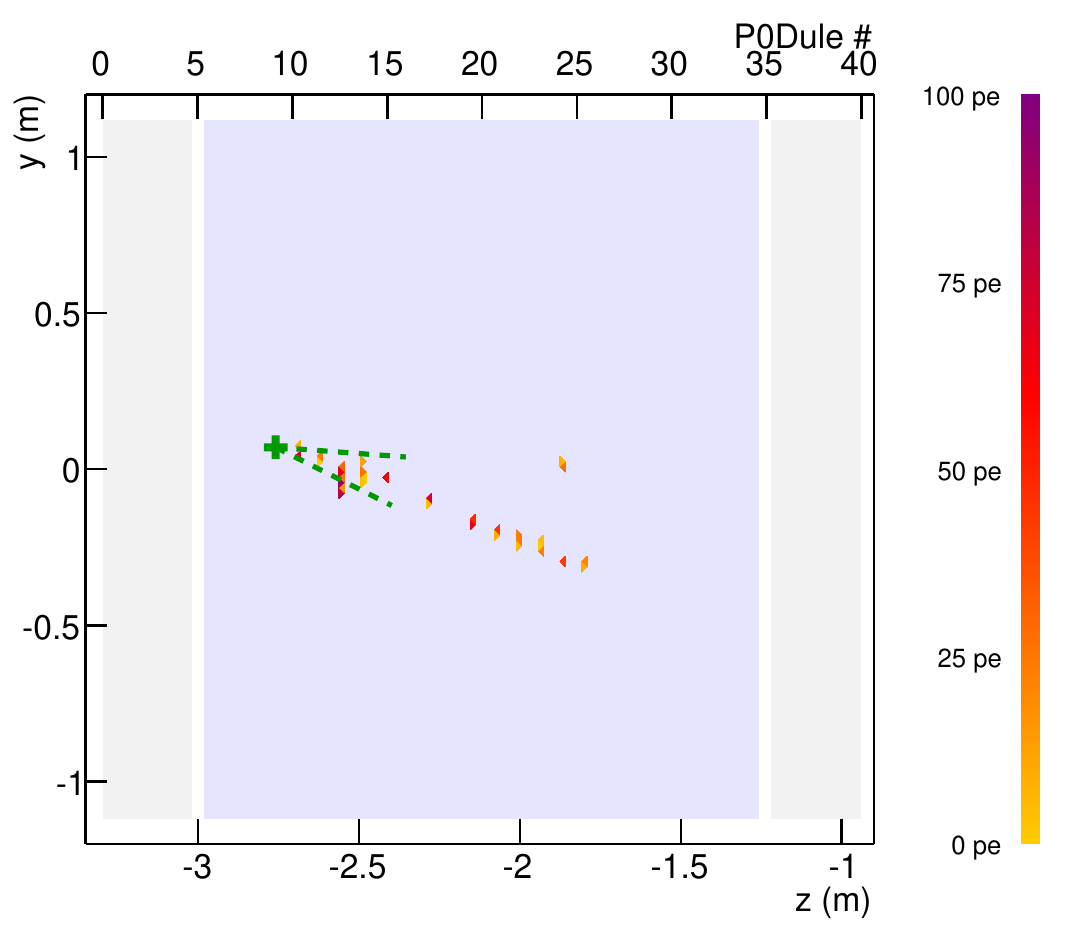}
	\includegraphics[width=\linewidth]{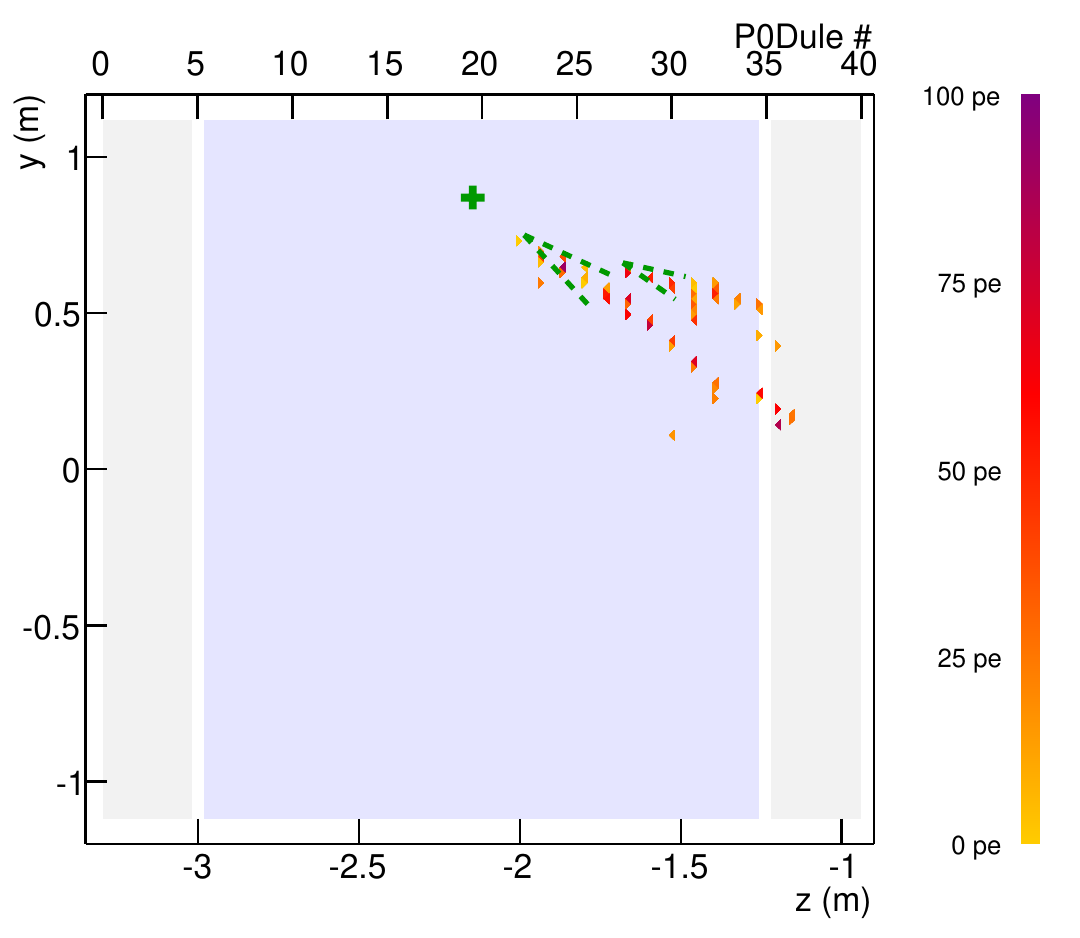}
	\caption{Side view of a \ccnue event (top) and a \pizero
          background event (bottom) reconstructed in the \podd.
          Triangles are hits colored by the charge deposited, the
          green cross symbol shows the reconstructed shower vertex,
          and the green dashed lines show the cones of reconstructed
          showers.}
	\label{fig:EventDisplay}
\end{figure}

The signal events for the analysis are the charged current $\nu_e$
interactions in the \podd{}.  A cut-based event selection using known
reconstruction characteristics was tuned to maximize the product of
efficiency and purity.  To avoid bias, the selection strategy was
developed based on Monte Carlo (MC) samples.  Event displays of a
typical \ccnue candidate and a \pizero background event selected in
the analysis are shown in Fig.~\ref{fig:EventDisplay}.

\subsection{\label{sec:SelectionCuts}Selection Cuts}

The event selection strategy focuses on identifying single high-energy
electron shower events with a vertex in the \podd.  As a
pre-selection, the reconstructed shower in the \podd must be in time
with the beam bunch time. The \podd reconstruction searches for both
tracks and showers with two independent algorithms, and the highest
energy track and the highest energy shower are used in the analysis.
The reconstruction algorithm builds tracks and showers from hits, but
as the shower reconstruction occurs after the track reconstruction the
algorighm needs to make sure that the hits shower reconstruction uses
are the same hits the track reconstruction uses, for each single
event. Therefore 80\% of the hits associated with the track and shower
are required to be the same.

In addition, events are selected where the angle of the reconstructed
shower with respect to the z-axis, which is approximately the beam
axis, is less than 45$^{\circ}$.  The scintillator bars of the \podd
have a triangular profile with angles of approximately 45$^{\circ}$.
Particles with an angle of more than 45$^{\circ}$ with respect to the
beam axis would therefore hit more than two adjacent bars in a layer.
The \podd reconstruction algorithm currently only handles up to two
adjacent bar hits in a layer, causing reconstruction failures for
higher angle tracks.

For this analysis, only events with a reconstructed neutrino energy of 1.5~GeV or more are
selected. Reconstructed neutrino energy is calculated from the reconstructed electron energy and the electron
angle using the quasi-elastic approximation. In this energy region, the majority of the $\nu_e$ flux
arises from kaon decays and the \podd shows good performance to
distinguish electrons from other particles.  In addition, using a high
neutrino energy cut improves the purity of the electron sample.

To reject muons, the median width of the selected track is used. In
each scintillator layer, the energy-weighted standard deviation of the
position of the hits reconstructed in the track is calculated as
follows:

\begin{enumerate}
	\item
	If the two hits with the highest deposited energy are in
        adjacent strips, replace them with a single hit. The new hit's
        position is at the energy-weighted average position of the two
        original hits, and its energy is the sum of the energies of
        the original hits. Any other hits in the layers are left
        unchanged. This procedure gives layers with minimum ionizing
        tracks very small (almost always zero) width.
	\item
	The energy-weighted standard deviation of the hit positions is
        calculated for each layer.
	\item
	Median width is the width of the middle layer after ordering
        by layer width.
\end{enumerate}

The design of the \podd with layers of high density materials (brass
and lead) causes electrons to shower.  The reconstructed track of an
electron is therefore typically wider than the reconstructed track of
a muon.  This feature can be used to distinguish muons and electrons
with the median width of the reconstructed candidate track.

\begin{figure}
	\includegraphics[width=\linewidth]{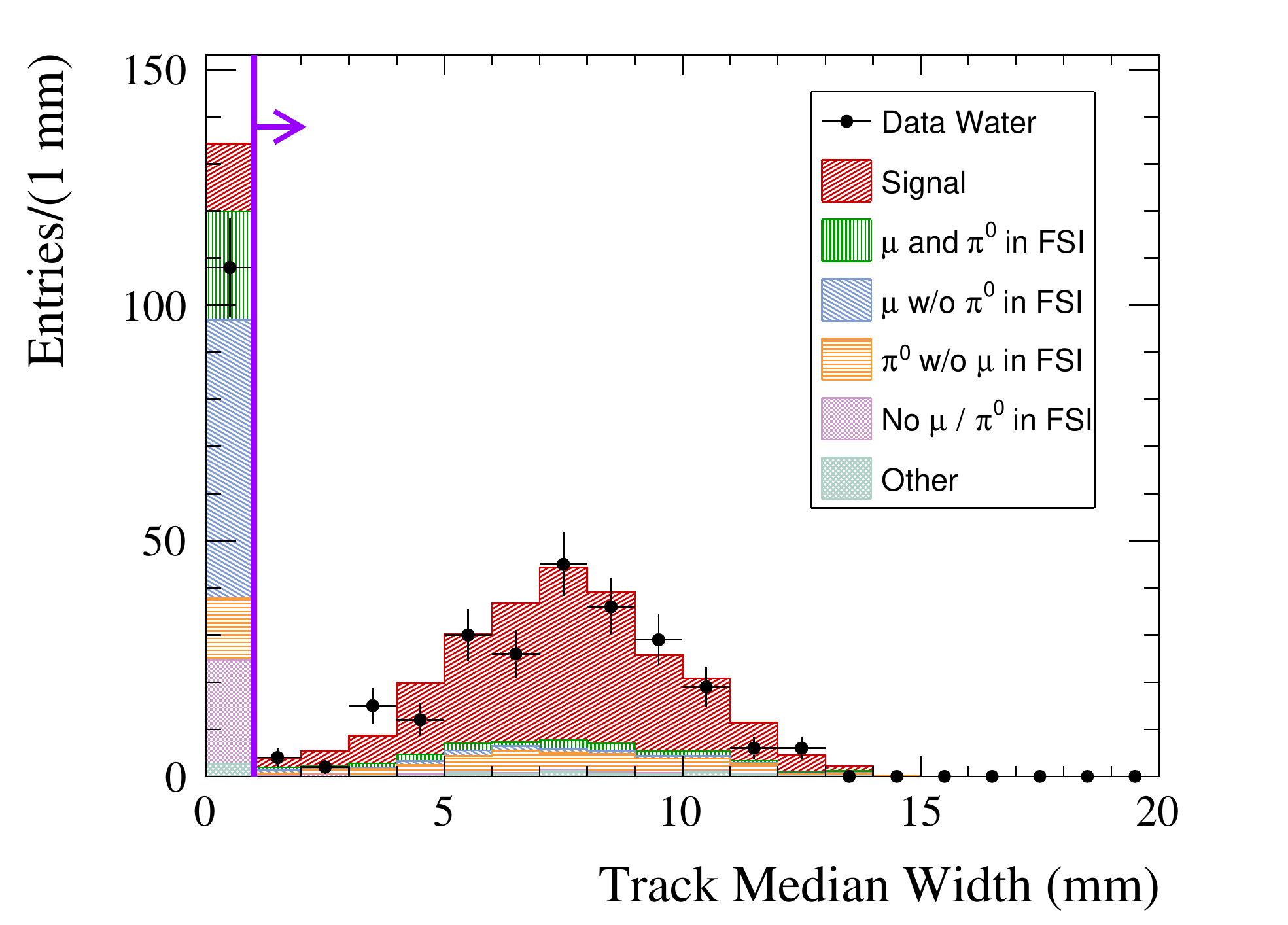}
	\includegraphics[width=\linewidth]{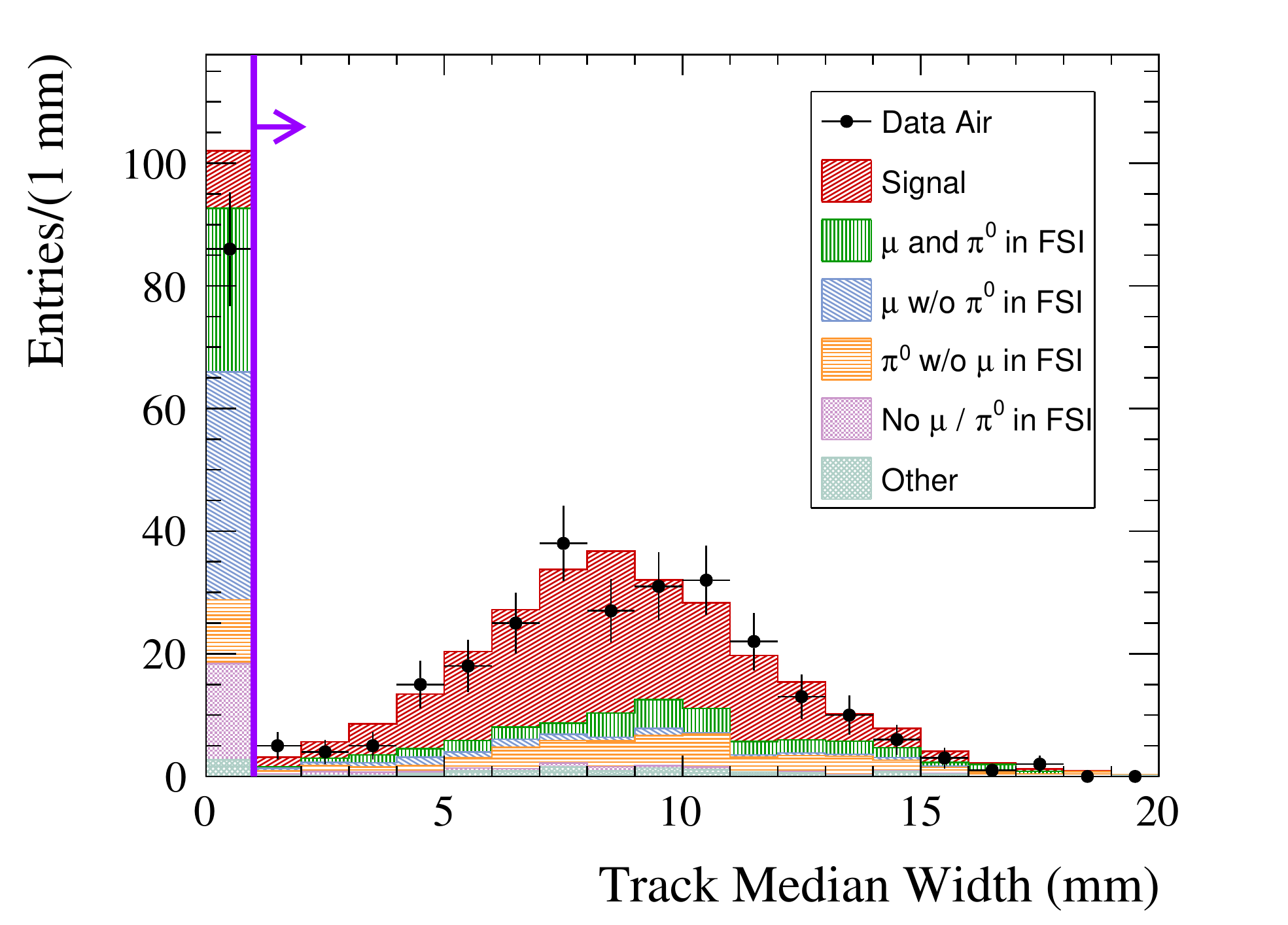}
	\caption{Distribution of events which pass all the selection
          criteria with the exception of the track median width cut,
          for water (top) and air configuration (bottom).  The vertical
          line shows the cut value used (1 mm). A sudden drop of events above 11~mm is an effect of shower median width cut.}
	\label{fig:nMinus1TrackMedianWidth}
\end{figure}

The track median width for events which pass all the selection criteria with the exception of the track median width cut, is shown in
Fig.~\ref{fig:nMinus1TrackMedianWidth} and indicates that most of the
background muon events are rejected by this cut.

\begin{figure}
	\includegraphics[width=\linewidth]{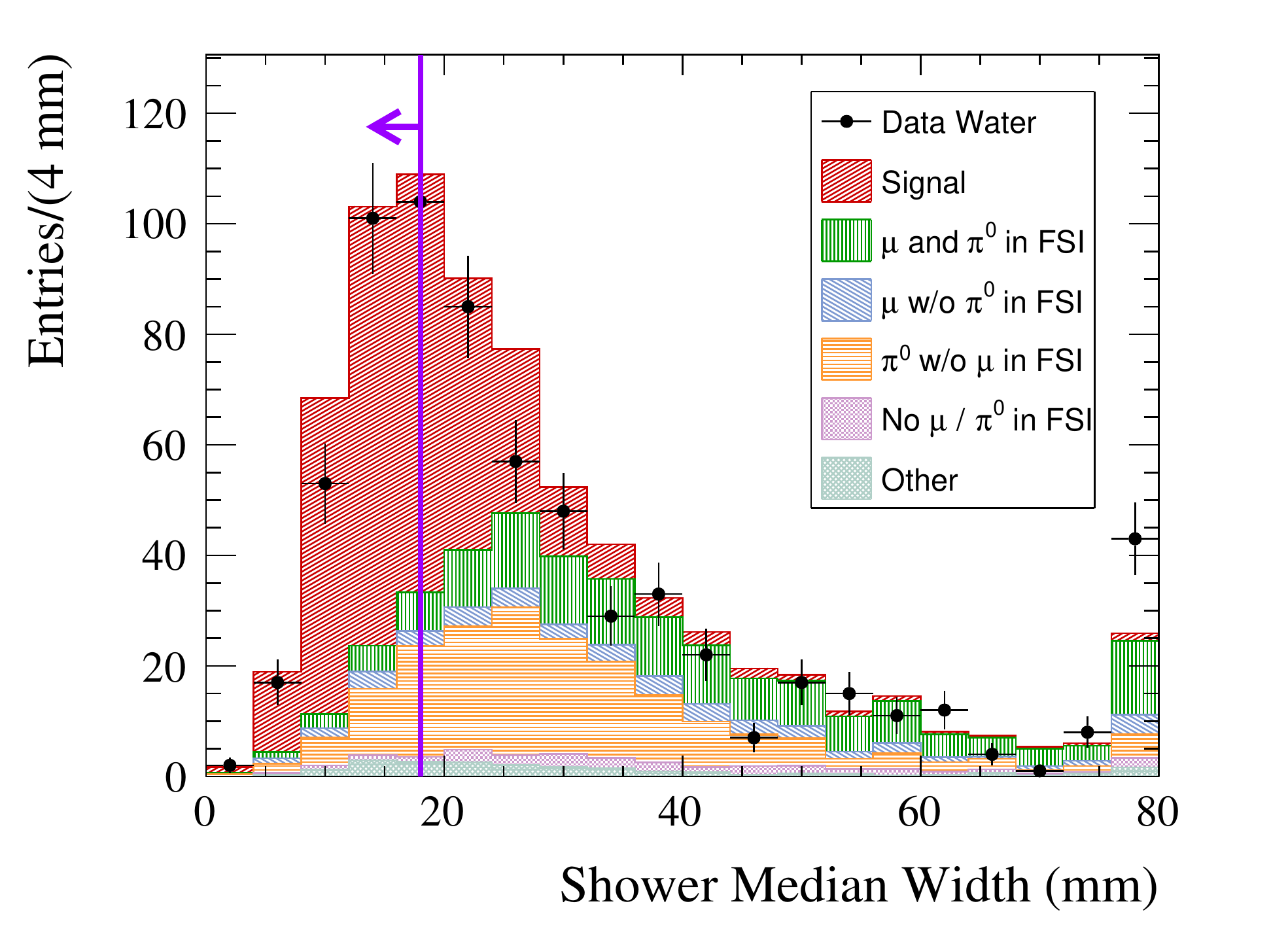}
	\includegraphics[width=\linewidth]{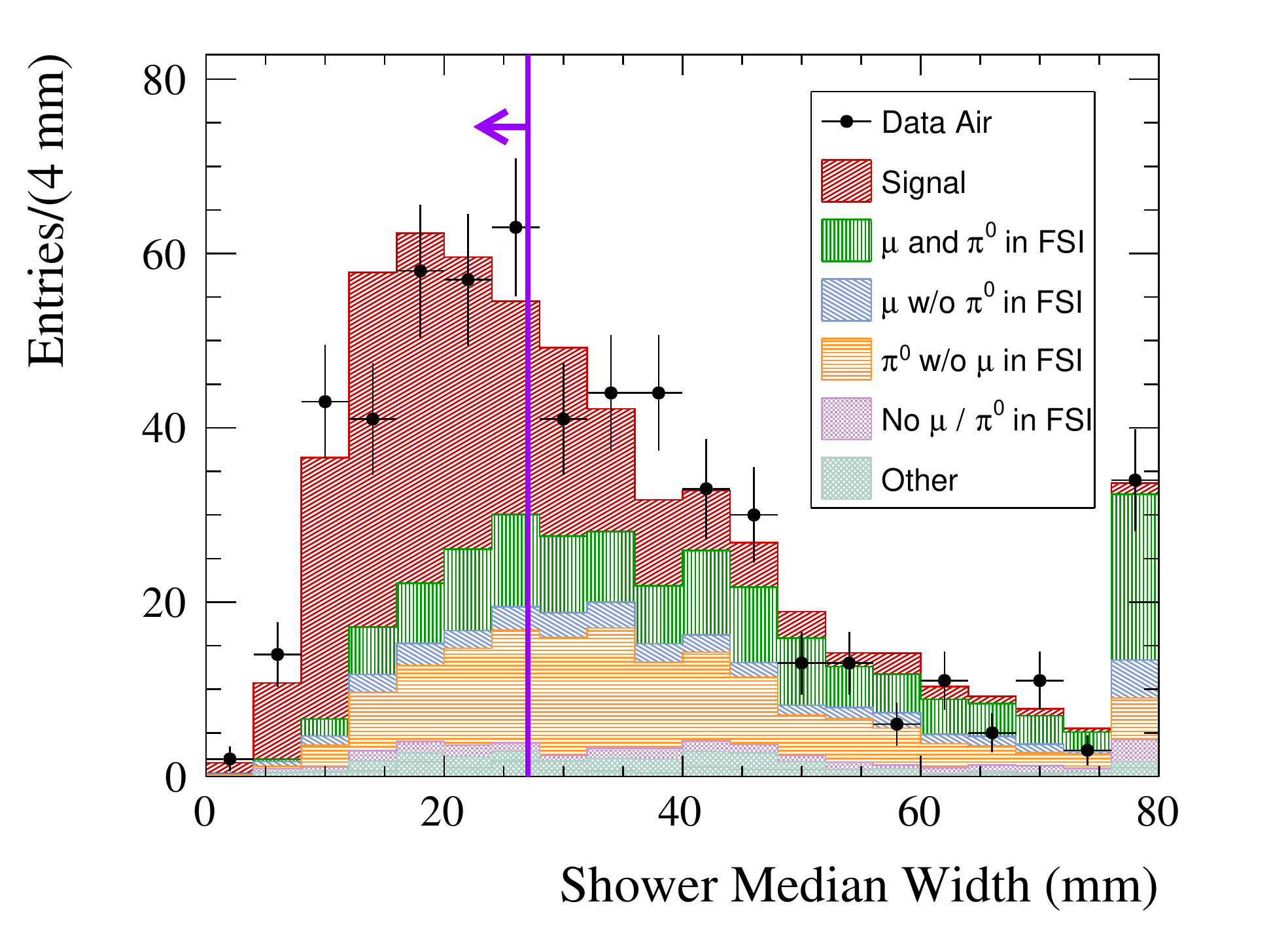}
	\caption{Distribution of events which pass all the selection
          criteria with the exception of the shower median width cut,
          for water (top) and air configuration (bottom).  The
          vertical line shows the applied cuts which are optimized for
          each configuration.}
	\label{fig:nMinus1ShowerMedianWidth}
\end{figure}

Similarly, to reject background events that contain neutral pions, a
cut is applied to the median width of the selected shower.  The shower
reconstruction looks for hits in a cone from the reconstructed vertex
position and combines them in one or more showers.  It can happen that
hits from several particles are combined in one reconstructed shower,
especially when they are almost overlapping.  The \podd $\nu_{e}$
analysis looks for events with a single electron.  Events with a very
wide candidate shower are rejected, because such events are more
likely background events with several particles.  The shower median
width is calculated the same way as the track median width.
Distributions of events which pass all the selection criteria with the
exception of the shower median width cut is shown in
Fig.~\ref{fig:nMinus1ShowerMedianWidth}. It shows many $\pi^0$
background events are rejected with this cut.

Finally, a cut is applied to the fraction of the event's charge that
is contained in the selected shower.  To select \ccnue events with a
high purity, the fraction of the event's charge contained in the
candidate shower is exactly 1.0 is required, which selects only events
with a single shower and without muon-like tracks in final state.

\subsection{\label{sec:SelectedEvents}Selected Event Samples}

The selected number of events passing all cuts predicted by the
simulation, both when the \podd is configured to contain water and
air, together with the number of selected data events are presented in
Table~\ref{tab:SelectedEvents}.  The water configuration simulation
events are separated into on-water and not-water events.  On-water
events are defined as events with true interaction vertex in the
water, and not-water events have the true interaction vertex on
scintillator, lead, brass, or other materials besides water.  All
events in the air configuration MC are not-water events as the water
targets are drained.

\begin{table}
\caption{The selected number of MC signal events, MC background
  events, and the total number of selected MC events normalized to
  data POT for water and air configuration are listed together with
  the selected data events.  In addition, the water configuration MC
  events are split up in on-water and not-water events. The errors
  correspond to the statistical uncertainty due to the limited MC
  statistics.}
\label{tab:SelectedEvents}
\begin{ruledtabular}
\begin{tabular}{lcccc}
      & MC Signal  & MC Background  & MC Total & Data  \\
      \hline
      Water                     &  $196.1 \pm 4.8$ & $56.7 \pm 2.7$ & $252.8 \pm 5.5$ & 230\\
      $\quad$ On-Water &  $60.2 \pm 2.6$ & $14.5 \pm 1.3$ & $74.7 \pm 2.9$ & \\
      $\quad$ Not-Water & $135.9 \pm 4.0$ & $42.2 \pm 2.3$ & $178.2 \pm 4.6$ & \\
      Air                          &  $173.6 \pm 4.6$ & $97.4 \pm 3.6$ & $271.0 \pm 5.8$ & 257\\
\end{tabular}
\end{ruledtabular}
\end{table}

\subsection{\label{sec:EfficiencyPurity}Efficiency and purity}

The efficiency $\epsilon$ and purity $p$ of the simulated electron
neutrino signal events, for water and air configurations, are
summarized in Table~\ref{tab:EfficiencyPurity}.  In the \podd water
configuration, events are split into events happening on water
(on-water) and events on scintillator, brass, and lead (not-water).


\begin{table}
 \caption{The signal efficiencies $\epsilon$ and purities $p$ are
   listed for water and air configuration.  Events of the \podd water
   configuration are split into events happening on-water and
   not-water.  The errors correspond to the statistical uncertainty
   due to the limited MC statistics. }
 \label{tab:EfficiencyPurity}
  \begin{ruledtabular}
 \begin{tabular}{lcc}
    & Efficiency $\epsilon$ & Purity $p$  \\
  \hline
  Water  & $(10.9 \pm 0.3)\%$ &   $(77.6 \pm 2.5)\%$ \\
    $\quad$ On-Water  & $(9.8 \pm 0.4)\%$ &   $(80.6 \pm 4.7)\%$\\
    $\quad$ Not-Water & $(11.5 \pm 0.4)\%$ &   $(76.3 \pm 3.0)\%$\\
   Air  & $(11.0 \pm 0.3)\%$ &   $(64.1 \pm 2.2)\%$\\
 \end{tabular}
\end{ruledtabular}
\end{table}

\begin{figure}
 \includegraphics[width=\linewidth]{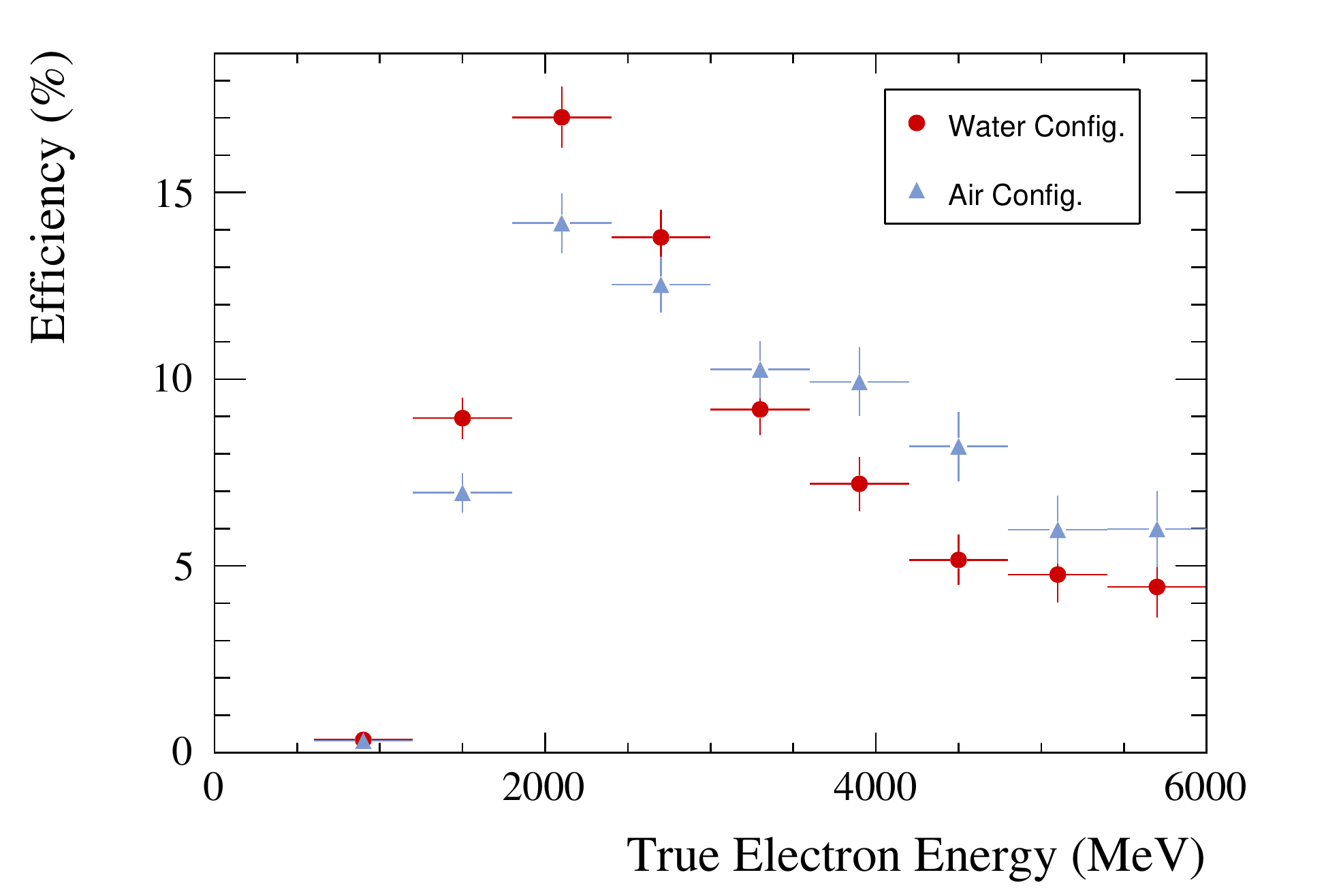}
 \caption{Selection efficiency of signal events as function of the
   true neutrino energy $E_{\text{true}}$ for water and air
   configuration.  The error bars correspond to the uncertainties due
   to limited MC statistics. }
 \label{fig:EfficiencyTrueNeutrinoEnergy}
\end{figure}

The selection efficiency of signal events as function of the true
neutrino energy $E_{\text{true}}$ for \podd water and air
configurations are shown in
Fig.~\ref{fig:EfficiencyTrueNeutrinoEnergy}.  The selection of low
energy signal events is suppressed by the high neutrino energy cut at
1.5~GeV while the selection of high energy signal events is suppressed
by the shower median width cut and the shower charge fraction cut.

\section{\label{sec:WaterSubtraction}Water Subtraction Method}

The measured $\nu_{e}$ interactions that were collected during \podd
water and air configuration running are compared with the number of
$\nu_{e}$ interactions predicted by the \podd water and air
configuration MC, respectively.  The measured number of $\nu_{e}$
interactions are extracted by subtracting the predicted MC background
$B$ from the selected data events $D$, resulting in:

\begin{eqnarray}
N_{\text{\ccnue{},water}}^{Data} &= D_{\text{water}}   - B_{\text{water}}\text{,\ and} \\[2mm]
N_{\text{\ccnue{},air}}^{Data} &= D_{\text{air}}  - B_{\text{air}}. 
\end{eqnarray}

The background subtracted data are then divided by the predicted Monte
Carlo signal $S$ to obtain the data/MC ratios for the water and air
configurations:

\begin{eqnarray}
R_{\text{water}} &= \frac{N_{\text{\ccnue{},water}}^{Data}}{S_{\text{water}}}\text{,\ and}\\[2mm]
R_{\text{air}} &= \frac{N_{\text{\ccnue{},air}}^{Data}}{S_{\text{air}}}.
\end{eqnarray}

To extract the measured number of on-water charged current $\nu_{e}$
interactions, the measured \ccnue interactions with \podd water and
air configurations are compared by taking into account the different
collected POT and the different reconstruction efficiencies for the
water and the air data sample using:

\begin{align}
N_{\text{\ccnue{},on-water}}^{Data} &= (D_{\text{water}} -
B_{\text{water}}) 
\notag \\[2mm]
&- \frac{\epsilon_{\text{not-water}} \cdot
  \text{POT}_{\text{water}}}{\epsilon_{\text{air}} \cdot
  \text{POT}_{\text{air}}} \cdot (D_{\text{air}} -  B_{\text{air}}).
\end{align}

In this formula, $\text{POT}_{\text{water}} = 2.64 \times 10^{20}$
($\text{POT}_{\text{air}} = 3.49 \times 10^{20}$) is the collected
data POT for the \podd water (air) configuration. The resulting
data/MC ratio for on-water \ccnue interactions is given by:

\begin{equation}
R_{\text{on-water}} = \frac{N_{\text{\ccnue{},on-water}} ^{Data}}{S_{\text{on-water}}}.
\end{equation}


\section{\label{sec:Systematics}Systematic Uncertainties}

The systematic uncertainties in the measurements are divided into
three categories: detector, reconstruction, and neutrino flux/cross
section uncertainties. Control sample events to study systematic effects in the measurement have been studied, but often the events in these control samples are not used for the final systematic uncertainty evaluation. The control sample events were found to be too similar to the signal events, or did not have the same background as the signal events.  For this reason a simple KS test is used for several of the systematic uncertainty tests, particularly where no deviation is indicated in the test.

\subsection{\label{sec:DetectorSystematics}Detector systematic
  uncertainties}

The detector's as-built mass and its mass in the Monte Carlo are
different.  The masses for water and air configurations as well as
different run periods also vary. These differences are incorporated in
the analysis procedure by re-weighting MC events with mass
uncertainties estimated to be 0.01 for all configurations.  Similarly,
the fiducial volume and the alignment of the \podd is
considered. Varying the fiducial volume by the MC vertex resolution
and shifting in \podd{} alignment provides an estimate of the
systematic uncertainties in data/MC ratios.  The uncertainties
obtained are smaller than 0.01 for all ratios making them negligible
in this measurement.

Possible systematic effects on the reconstructed electron energy are
also studied.  The effects are investigated by changing the
reconstructed energy scale to observe the differences in \ccnue
data/MC ratios.  The possible effects are as follows: 1. \podd
material density and thickness, 2. drifts in the \podd response over
time, and 3. the simulation (GEANT4) uncertainty in the electron
energy deposition.  It is assumed the water and air configuration are
correlated for the \podd material density and thickness only. The
resulting systematic uncertainties for water ($R_{\text{water}}$), air
($R_{\text{air}}$), and on-water ($R_{\text{on-water}}$) are 0.05,
0.05, and 0.10 respectively.

\subsection{\label{sec:ReconSystematics}Reconstruction systematic uncertainties}

\subsubsection{\label{sec:TrackPID}Track PID}

As described earlier at the beginning of Section
\ref{sec:EventSelection}, the classification of the reconstructed
tracks is based on the \podd PID.  Differences in the PID between data
and MC can therefore cause systematic uncertainties in the \ccnue
data/MC ratios.

A PID study with stopping muons in the \podd was performed to estimate
this uncertainty, and a map of mis-PID between a data sample and a
simulation of stopping muons was constructed.  To estimate the impact
of the track PID uncertainty on the \ccnue data/MC ratios, the MC
signal and background was weighted according to the uncertainty of the
map.  The systematic parameter values were randomly varied assuming
that the water and air samples are uncorrelated and also that the
signal and background uncertainties are uncorrelated.  The
uncertainties for water ($R_{\text{water}}$), air ($R_{\text{air}}$),
and on-water ($R_{\text{on-water}}$) were determined to be 0.05, 0.05,
and 0.09 respectively.

\subsubsection{\label{sec:MedianWidth}Track and Shower Median width}

To estimate the systematic uncertainty caused by the track median
width, the plots with all selection criteria applied but failing the
track median width cut (the N-1 plots) are integrated, and a
Kolmogorov-Smirnov test is performed to test if the data and the Monte
Carlo event distributions are consistent\cite{Kolmogorov,Smirnov}.
The Kolmogorov-Smirnov test returns a p-value of 91.2\% for water and
92.2\% for air configuration indicating that there are no significant
evidence for a shift between the data and MC event distributions.  The
systematic uncertainty due to the track median width cut is therefore
negligible for this analysis.

The threshold of the shower median width cut is placed in a region
with a large number of events.  The systematic uncertainty on the
measured shower median width therefore has a larger impact on the
\ccnue data/MC ratios than the track median width uncertainty does.
To estimate the systematic uncertainty caused by the shower median
width, the N-1 plots are integrated, and a Kolmogorov-Smirnov test is
performed.  The Kolmogorov-Smirnov test returns a p-value of 50.0\%
for water and 65.9\% for air configuration.  To determine a reasonable
scaling factor range for Monte Carlo, different scaling factors from
0.9 to 1.1 were applied to Monte Carlo and the resulting p-values were
studied.  For a p-value of 68\%, the peak scaling factor ranged from
0.98 to 1.02.  The systematic effect on the \ccnue data/MC ratios for
$R_{\text{water}}$, $R_{\text{air}}$, and $R_{\text{on-water}}$ coming
from the shower median width are estimated by varying the scaling
factor that is applied to the MC shower median width.  The
uncertainties obtained for $R_{\text{water}}$, $R_{\text{air}}$, and
$R_{\text{on-water}}$ are 0.04, 0.04, and 0.08 respectively.

\subsubsection{\label{sec:ChargeFraction}Shower Charge Fraction}

To estimate the possible impact of systematic effects of the shower
charge fraction on the analysis, additional reconstructed objects with
low energy are studied.  Such additional tracks or showers would cause
an event to fail the shower charge fraction selection criteria.
Looking at the event distribution of these events, the only hint for a
systematic difference between data and MC appears in the highest bin
of the air configuration.  Events with a shower charge fraction
between 0.98 and 1.00 which pass all other selection criteria are
analyzed to estimate the systematic uncertainty.  The data/MC
difference in this region is considered to be the uncertainty on the
MC events in the signal region, resulting in the systematic
uncertainties for $R_{\text{water}}$, $R_{\text{air}}$, and
$R_{\text{on-water}}$ of 0.01, 0.04, and 0.04 respectively.

\subsection{\label{sec:FluxSystematics}Flux and cross section systematic uncertainties}

For the inclusion of the flux and cross section systematic
uncertainties in the analysis, each analyzed MC event is re-weighted
according to the uncertainties of the flux and cross section
parameters which are correlated.  The parameter values and
uncertainties are provided by different external measurements such as
NA61 and other hadronic production experiments, and these parameters
are then fitted to ND280 data from TPC and FGD, the other subdetectors
of ND280 than \podd. The systematic parameters and their uncertainties
obtained from the fit to the ND280 data, which includes 25 flux
parameters, 6 FSI parameters, 2 NEUT parameters, and 13 neutrino
interaction parameters, has been studied in
Ref. \cite{PhysRevD.89.092003}.
 
To obtain the flux and cross section systematic uncertainties, the
systematic parameters are thrown according to the covariance matrix
and the analysis described in Section~\ref{sec:WaterSubtraction} is
then applied to each throw.  The distributions are fit with single
Gaussians and the resulting width is considered to be the flux and
cross section systematic uncertainty for the analysis.  The
uncertainties obtained for water ($R_{\text{water}}$), air
($R_{\text{air}}$), and on-water ($R_{\text{on-water}}$) are 0.07,
0.09, and 0.06 respectively.

\subsection{\label{sec:SystematicsSummary}Summary of the systematic uncertainties}

All systematic uncertainties on the \ccnue data/MC ratios for water
($R_{\text{water}}$), air ($R_{\text{air}}$), and on-water
($R_{\text{on-water}}$) that were estimated in the previous sections
are summarized in Table~\ref{tab:SummarySystematics}.  This table also
shows the total systematic uncertainty.

\begin{table}
 \caption{Summary of systematic uncertainties on the \ccnue data/MC ratios for water ($R_{\text{water}}$), air ($R_{\text{air}}$), and on-water ($R_{\text{on-water}}$). }
 \label{tab:SummarySystematics}
\begin{ruledtabular}
 \begin{tabular}{lccc}
    Systematic Uncertainty & $R_{\text{water}}$ & $R_{\text{air}}$ & $R_{\text{on-water}}$\\

  \hline
      MC Statistics & 0.03  & 0.04 & 0.12  \\
   \hline
   \podd Mass & 0.01 & 0.01 & 0.01 \\
    \podd Fiducial Volume & $<0.01$ & $<0.01$ & $<0.01$ \\
    \podd Alignment & $<0.01$ & $<0.01$ & $<0.01$ \\
    \hline
    Energy Scale & 0.05 & 0.05 & 0.10 \\
    \hline
    Hit Matching & $<0.01$ & $<0.01$ & $<0.01$\\
    Track PID & 0.05 & 0.05 & 0.09 \\
    Energy Resolution & $<0.01$ & $<0.01$ & 0.01 \\
    Angular Resolution & $<0.01$ & $<0.01$ & 0.01\\
    Track Median Width & $<0.01$ & $<0.01$ & $<0.01$\\
    Shower Median Width & 0.04 & 0.04 & 0.08\\
    Shower Charge Fraction & 0.01 & 0.04 & 0.04\\
    \hline
    Flux and Cross Sections & 0.07 & 0.09 & 0.06 \\
    \hline
    Total & 0.11 & 0.13 & 0.21 \\
 \end{tabular}
 \end{ruledtabular}
 \end{table}

\section{\label{sec:Results}Results}

The results obtained for the background subtracted data/MC ratio ($R$)
for water configuration, air configuration, and on-water are:

\begin{eqnarray}
R_{\text{water}} &= 0.89 \pm 0.08  \text{ (stat.)} \pm 0.11  \text{ (sys.),} \\[2mm]
R_{\text{air}} &= 0.90 \pm 0.09  \text{ (stat.)} \pm 0.13  \text{ (sys.),\ and} \\[2mm]
R_{\text{on-water}} &= 0.87 \pm 0.33  \text{ (stat.)} \pm 0.21  \text{
  (sys.).} 
\end{eqnarray}

The ratios are consistent with 1, within statistical and systematic
uncertainties. For the on-water ratio, uncertainties are relatively
large due to limited statistics and the impact of the subtraction
method.

\begin{figure}[!ht]
	\includegraphics[width=0.51\textwidth]{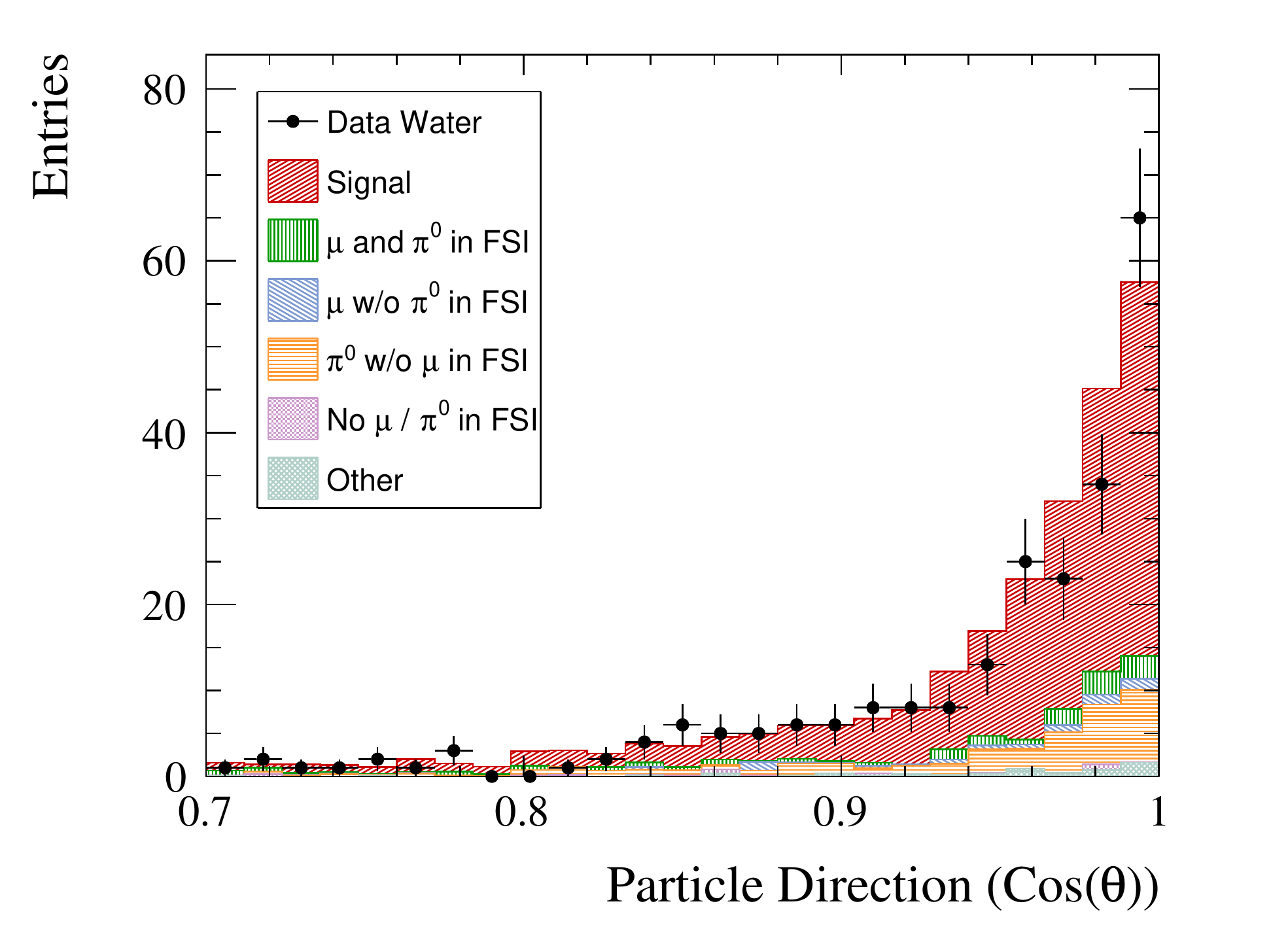}
	\includegraphics[width=0.51\textwidth]{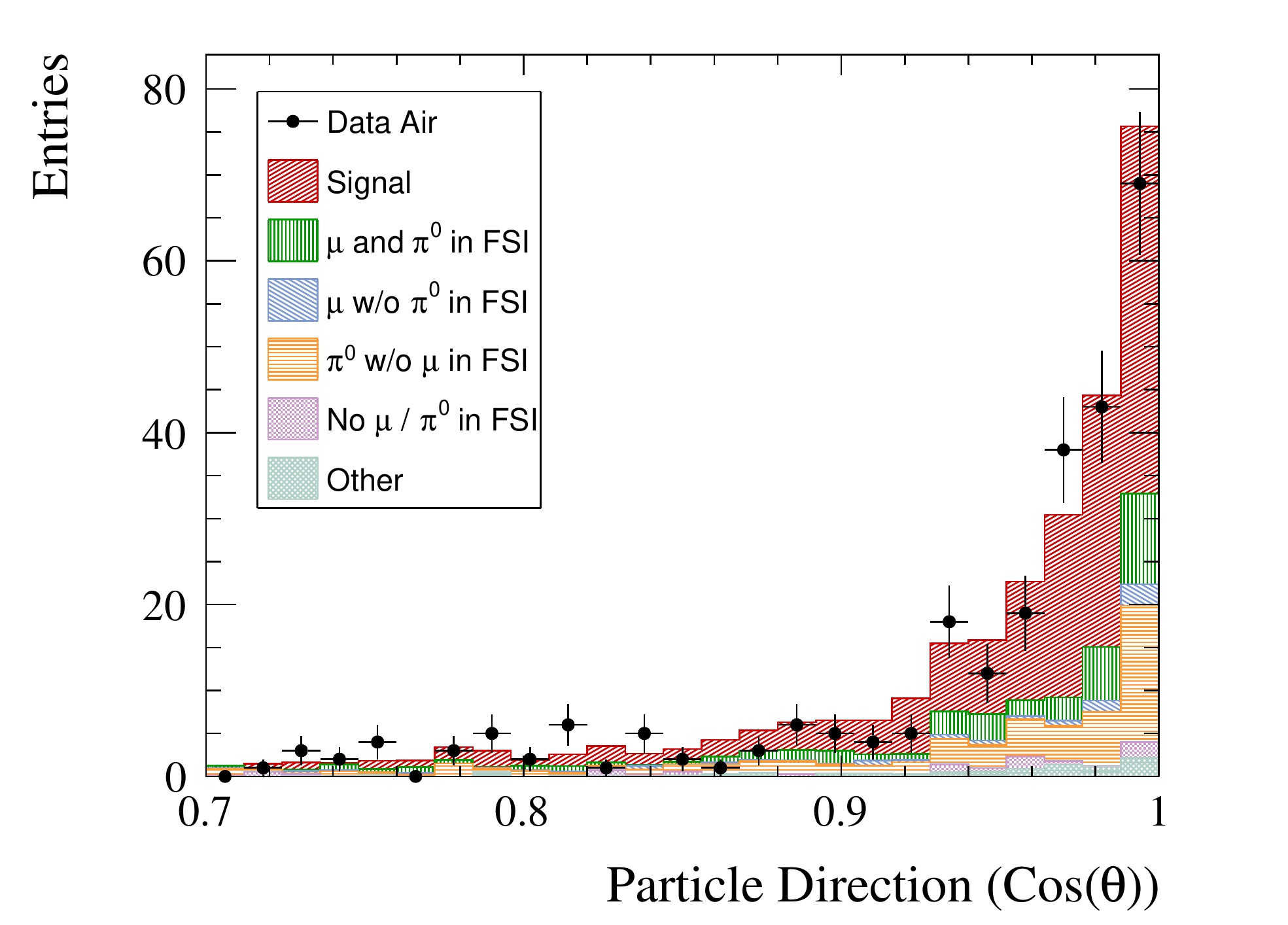}
	\caption{Events passing the event selection as a function of
          the particle direction for water (top) and air configuration
          (bottom). The MC events are normalized to data POT, and the
          fit results from ND280 are applied.}
	\label{fig:finalPlotsDirection}
\end{figure}

\begin{figure}[ht!]
	\includegraphics[width=0.51\textwidth]{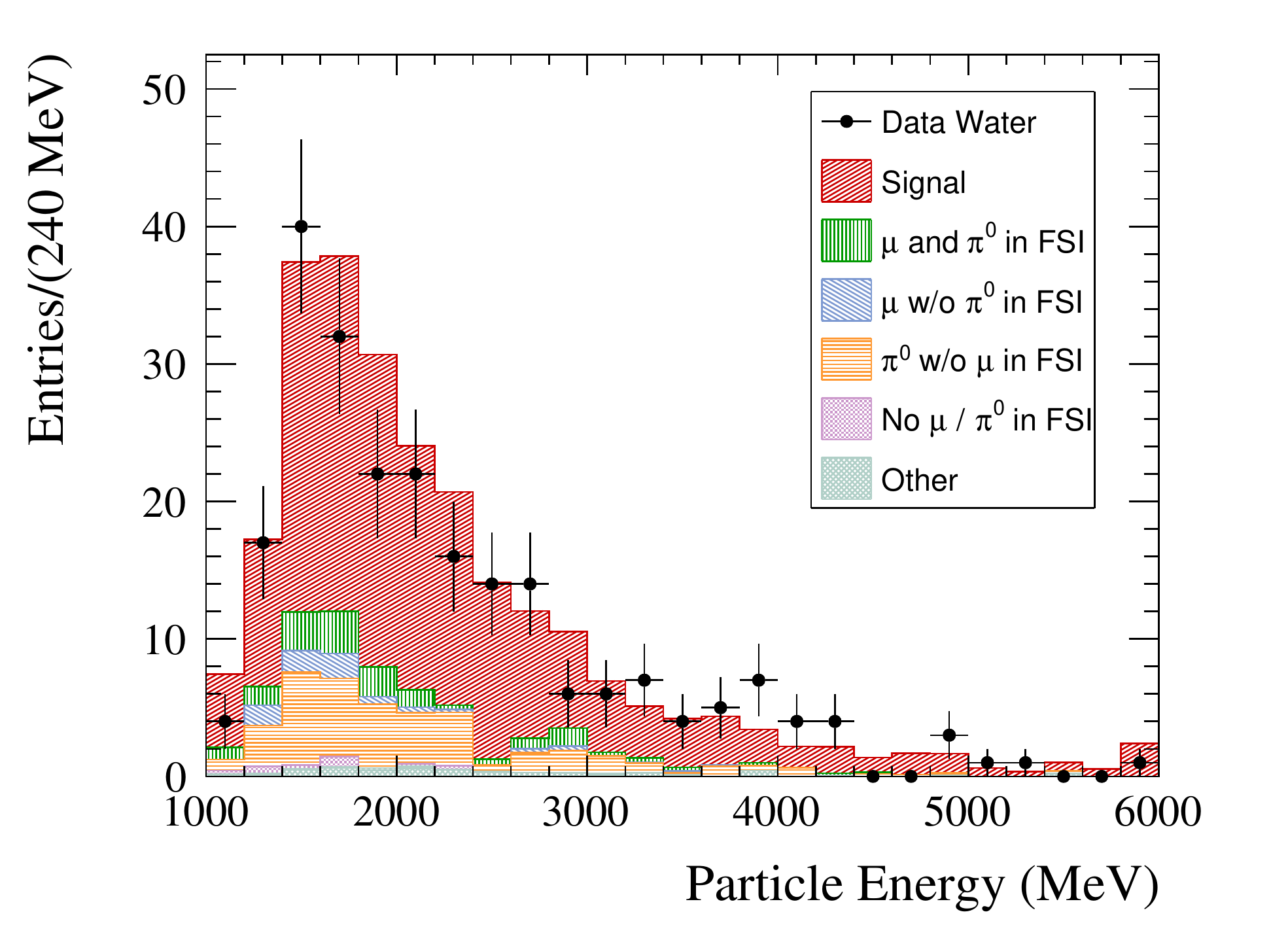}
	\includegraphics[width=0.51\textwidth]{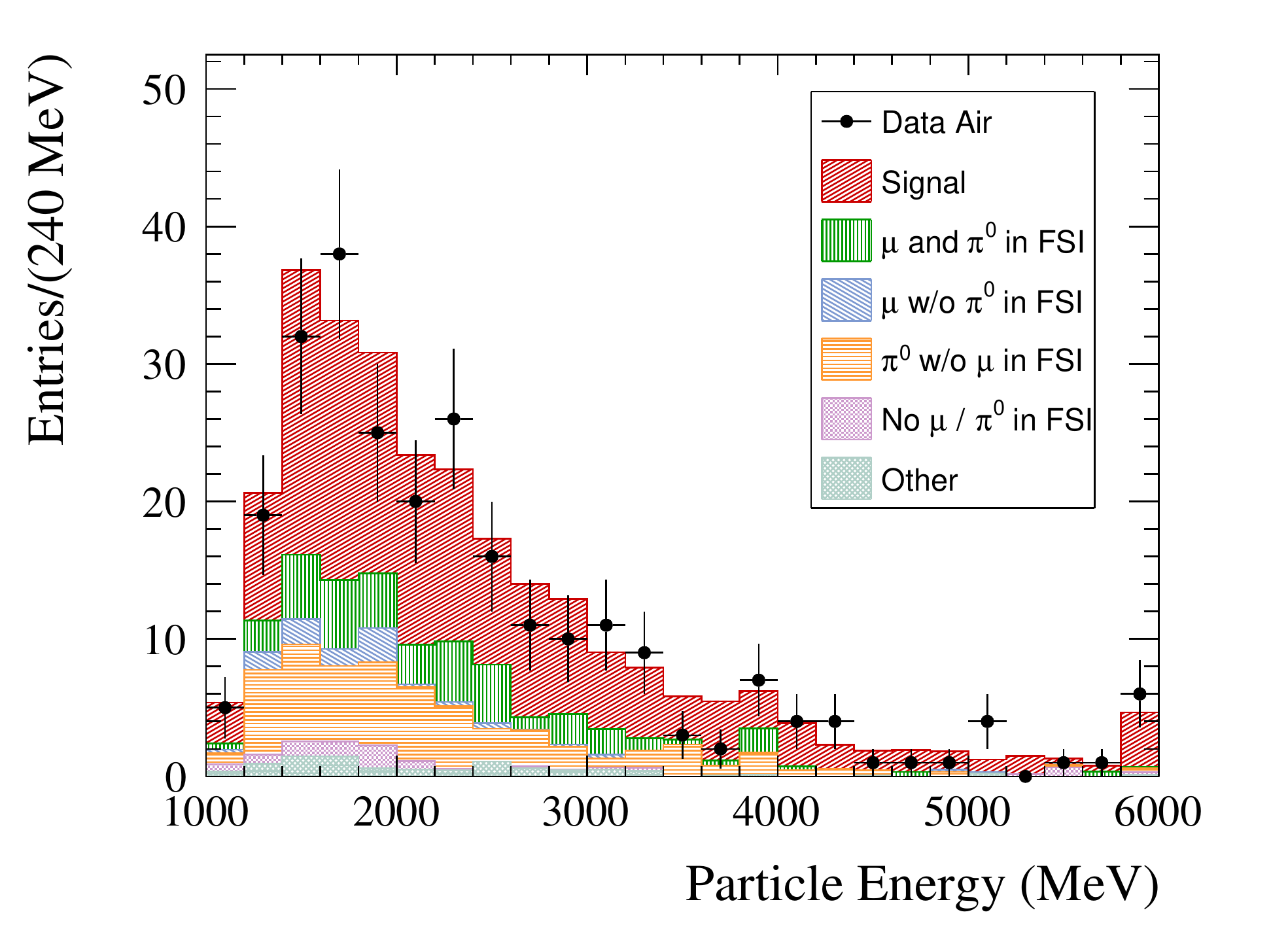}
	\caption{Events passing the event selection as a function of
          the particle energy for water (top) and air configuration
          (bottom). The MC events are normalized to data POT, and the
          fit results from ND280 are applied.}

	\label{fig:finalPlotsEnergy}
\end{figure}

For the selected events, the distribution of the reconstructed
particle directions is shown in Fig.~\ref{fig:finalPlotsDirection} and
the distribution of particle energies is shown in
Fig.~\ref{fig:finalPlotsEnergy}. This result indicates that the beam
$\nu_e$ component in high energy region measured in the data is
consistent with expectations after including constraints from the
ND280 data for all configurations.

\section{\label{sec:Conclusion}Conclusion}

In conclusion, measurements of \ccnue{} interactions using the ND280
\podd{} have been made. The \podd{} includes fillable water targets
which allows separate measurements for the water and air
configurations of the ND280 \podd{} as well as the measurement of
$\nu_e$ on-water interactions above 1.5 GeV in a predominantly
$\nu_\mu$ beam. About $\sim$85\% of the selected sample comes from the
decay of kaons.

The 230 (257) water configuration (air configuration) electron
neutrino candidate events selected in the data are in good agreement
with the prediction for the water configuration, the air
configuration, and for the on-water subtraction samples
respectively. The measurement is statistically limited, especially for
on-water, but it will be improved in the future, since collection of
ten times more data is planned in the coming years.  Furthermore,
studies and improvements to the reconstruction algorithms are being
investigated to lower the energy threshold, which will lead to the
measurement of the $\nu_e$ cross section on water.

This is the first $\nu_e$ interaction rate measurement on water in the
few GeV energy region.  Interactions of $\nu_e$ on water are of
particular interest for long-baseline neutrino oscillation
experiments, and atmospheric neutrino experiments using water
Cherenkov detectors with the aim to measure $CP$ violation in the
lepton sector.

\FloatBarrier

\begin{acknowledgments}

We thank the J-PARC staff for superb accelerator performance and the
CERN NA61 collaboration for providing valuable particle production
data. We acknowledge the support of MEXT, Japan; NSERC, NRC and CFI,
Canada; CEA and CNRS/IN2P3, France; DFG, Germany; INFN, Italy;
National Science Centre (NCN), Poland; RSF, RFBR and MES, Russia;
MINECO and ERDF funds, Spain; SNSF and SER, Switzerland; STFC, UK; and
DOE, USA. We also thank CERN for the UA1/NOMAD magnet, DESY for the
HERA-B magnet mover system, NII for SINET4, the WestGrid and SciNet
consortia in Compute Canada, GridPP, UK. In addition participation of
individual researchers and institutions has been further supported by
funds from: ERC (FP7), EU; JSPS, Japan; Royal Society, UK; DOE Early
Career program, USA.

\end{acknowledgments}

\bibliography{P0DNuE}

\providecommand{\noopsort}[1]{}\providecommand{\singleletter}[1]{#1}%
\begin{thebibliography}{29}%
\makeatletter
\providecommand \@ifxundefined [1]{%
 \@ifx{#1\undefined}
}%
\providecommand \@ifnum [1]{%
 \ifnum #1\expandafter \@firstoftwo
 \else \expandafter \@secondoftwo
 \fi
}%
\providecommand \@ifx [1]{%
 \ifx #1\expandafter \@firstoftwo
 \else \expandafter \@secondoftwo
 \fi
}%
\providecommand \natexlab [1]{#1}%
\providecommand \enquote  [1]{``#1''}%
\providecommand \bibnamefont  [1]{#1}%
\providecommand \bibfnamefont [1]{#1}%
\providecommand \citenamefont [1]{#1}%
\providecommand \href@noop [0]{\@secondoftwo}%
\providecommand \href [0]{\begingroup \@sanitize@url \@href}%
\providecommand \@href[1]{\@@startlink{#1}\@@href}%
\providecommand \@@href[1]{\endgroup#1\@@endlink}%
\providecommand \@sanitize@url [0]{\catcode `\\12\catcode `\$12\catcode
  `\&12\catcode `\#12\catcode `\^12\catcode `\_12\catcode `\%12\relax}%
\providecommand \@@startlink[1]{}%
\providecommand \@@endlink[0]{}%
\providecommand \url  [0]{\begingroup\@sanitize@url \@url }%
\providecommand \@url [1]{\endgroup\@href {#1}{\urlprefix }}%
\providecommand \urlprefix  [0]{URL }%
\providecommand \Eprint [0]{\href }%
\providecommand \doibase [0]{http://dx.doi.org/}%
\providecommand \selectlanguage [0]{\@gobble}%
\providecommand \bibinfo  [0]{\@secondoftwo}%
\providecommand \bibfield  [0]{\@secondoftwo}%
\providecommand \translation [1]{[#1]}%
\providecommand \BibitemOpen [0]{}%
\providecommand \bibitemStop [0]{}%
\providecommand \bibitemNoStop [0]{.\EOS\space}%
\providecommand \EOS [0]{\spacefactor3000\relax}%
\providecommand \BibitemShut  [1]{\csname bibitem#1\endcsname}%
\let\auto@bib@innerbib\@empty
\bibitem [{\citenamefont {Hayato}(2002)}]{NEUT}%
  \BibitemOpen
  \bibfield  {author} {\bibinfo {author} {\bibfnamefont {Y.}~\bibnamefont
  {Hayato}},\ }\href {\doibase DOI: 10.1016/S0920-5632(02)01759-0} {\bibfield
  {journal} {\bibinfo  {journal} {{Nuclear Physics B - Proceedings
  Supplements}}\ }\textbf {\bibinfo {volume} {112}},\ \bibinfo {pages} {171 }
  (\bibinfo {year} {2002})}\BibitemShut {NoStop}%
\bibitem [{\citenamefont {Blietschau}\ \emph {et~al.}(1978)\citenamefont
  {Blietschau} \emph {et~al.}}]{Blietschau:1978mu}%
  \BibitemOpen
  \bibfield  {author} {\bibinfo {author} {\bibfnamefont {J.}~\bibnamefont
  {Blietschau}} \emph {et~al.} (\bibinfo {collaboration} {Gargamelle}),\
  }\href@noop {} {\bibfield  {journal} {\bibinfo  {journal} {Nucl. Phys.}\
  }\textbf {\bibinfo {volume} {B133}},\ \bibinfo {pages} {205} (\bibinfo {year}
  {1978})}\BibitemShut {NoStop}%
\bibitem [{\citenamefont {Auerbach}\ \emph {et~al.}(2001)\citenamefont
  {Auerbach} \emph {et~al.}}]{Auerbach:2001wg}%
  \BibitemOpen
  \bibfield  {author} {\bibinfo {author} {\bibfnamefont {L.~B.}\ \bibnamefont
  {Auerbach}} \emph {et~al.} (\bibinfo {collaboration} {LSND}),\ }\href@noop {}
  {\bibfield  {journal} {\bibinfo  {journal} {Phys. Rev.D}\ }\textbf {\bibinfo
  {volume} {63}},\ \bibinfo {pages} {112001} (\bibinfo {year} {2001})},\
  \Eprint {http://arxiv.org/abs/hep-ex/0101039} {arXiv:hep-ex/0101039}
  \BibitemShut {NoStop}%
\bibitem [{\citenamefont {Formaggio}\ and\ \citenamefont
  {Zeller}(2012)}]{RevModPhys.84.1307}%
  \BibitemOpen
  \bibfield  {author} {\bibinfo {author} {\bibfnamefont {J.~A.}\ \bibnamefont
  {Formaggio}}\ and\ \bibinfo {author} {\bibfnamefont {G.~P.}\ \bibnamefont
  {Zeller}},\ }\href {\doibase 10.1103/RevModPhys.84.1307} {\bibfield
  {journal} {\bibinfo  {journal} {Rev. Mod. Phys.}\ }\textbf {\bibinfo {volume}
  {84}},\ \bibinfo {pages} {1307} (\bibinfo {year} {2012})}\BibitemShut
  {NoStop}%
\bibitem [{\citenamefont {Abe}\ \emph {et~al.}(2011{\natexlab{a}})\citenamefont
  {Abe} \emph {et~al.}}]{Abe:2011ks}%
  \BibitemOpen
  \bibfield  {author} {\bibinfo {author} {\bibfnamefont {K.}~\bibnamefont
  {Abe}} \emph {et~al.} (\bibinfo {collaboration} {T2K Collaboration}),\ }\href
  {\doibase 10.1016/j.nima.2011.06.067} {\bibfield  {journal} {\bibinfo
  {journal} {Nucl.Instrum.Meth.}\ }\textbf {\bibinfo {volume} {A659}},\
  \bibinfo {pages} {106} (\bibinfo {year} {2011}{\natexlab{a}})},\ \Eprint
  {http://arxiv.org/abs/1106.1238} {arXiv:1106.1238 [physics.ins-det]}
  \BibitemShut {NoStop}%
\bibitem [{\citenamefont {Arafune}\ \emph {et~al.}(1997)\citenamefont {Arafune}
  \emph {et~al.}}]{PhysRevD.56.3093}%
  \BibitemOpen
  \bibfield  {author} {\bibinfo {author} {\bibfnamefont {J.}~\bibnamefont
  {Arafune}} \emph {et~al.},\ }\href {\doibase 10.1103/PhysRevD.56.3093}
  {\bibfield  {journal} {\bibinfo  {journal} {Phys. Rev. D}\ }\textbf {\bibinfo
  {volume} {56}},\ \bibinfo {pages} {3093} (\bibinfo {year}
  {1997})}\BibitemShut {NoStop}%
\bibitem [{\citenamefont {Abe}\ \emph {et~al.}(2014{\natexlab{a}})\citenamefont
  {Abe} \emph {et~al.}}]{Abe:2013hdq}%
  \BibitemOpen
  \bibfield  {author} {\bibinfo {author} {\bibfnamefont {K.}~\bibnamefont
  {Abe}} \emph {et~al.} (\bibinfo {collaboration} {T2K Collaboration}),\ }\href
  {\doibase 10.1103/PhysRevLett.112.061802} {\bibfield  {journal} {\bibinfo
  {journal} {Phys.Rev.Lett.}\ }\textbf {\bibinfo {volume} {112}},\ \bibinfo
  {pages} {061802} (\bibinfo {year} {2014}{\natexlab{a}})},\ \Eprint
  {http://arxiv.org/abs/1311.4750} {arXiv:1311.4750 [hep-ex]} \BibitemShut
  {NoStop}%
\bibitem [{\citenamefont {Abe}\ \emph {et~al.}(2013)\citenamefont {Abe} \emph
  {et~al.}}]{PhysRevD.87.012001}%
  \BibitemOpen
  \bibfield  {author} {\bibinfo {author} {\bibfnamefont {K.}~\bibnamefont
  {Abe}} \emph {et~al.} (\bibinfo {collaboration} {T2K Collaboration}),\ }\href
  {\doibase 10.1103/PhysRevD.87.012001} {\bibfield  {journal} {\bibinfo
  {journal} {Phys. Rev. D}\ }\textbf {\bibinfo {volume} {87}},\ \bibinfo
  {pages} {012001} (\bibinfo {year} {2013})}\BibitemShut {NoStop}%
\bibitem [{\citenamefont {Abe}\ \emph {et~al.}(2011{\natexlab{b}})\citenamefont
  {Abe} \emph {et~al.}}]{Abe:2011ts}%
  \BibitemOpen
  \bibfield  {author} {\bibinfo {author} {\bibfnamefont {K.}~\bibnamefont
  {Abe}} \emph {et~al.},\ }\href@noop {} {\enquote {\bibinfo {title} {{Letter
  of Intent: The Hyper-Kamiokande Experiment, Detector Design and Physics
  Potential}},}\ } (\bibinfo {year} {2011}{\natexlab{b}}),\ \Eprint
  {http://arxiv.org/abs/1109.3262} {arXiv:1109.3262 [hep-ex]} \BibitemShut
  {NoStop}%
\bibitem [{\citenamefont {Adams}\ \emph {et~al.}(2013)\citenamefont {Adams}
  \emph {et~al.}}]{Adams:2013qkq}%
  \BibitemOpen
  \bibfield  {author} {\bibinfo {author} {\bibfnamefont {C.}~\bibnamefont
  {Adams}} \emph {et~al.} (\bibinfo {collaboration} {LBNE Collaboration}),\
  }\href@noop {} {\enquote {\bibinfo {title} {{Scientific Opportunities with
  the Long-Baseline Neutrino Experiment}},}\ } (\bibinfo {year} {2013}),\
  \Eprint {http://arxiv.org/abs/1307.7335} {arXiv:1307.7335 [hep-ex]}
  \BibitemShut {NoStop}%
\bibitem [{\citenamefont {Stahl}\ \emph {et~al.}(2012)\citenamefont {Stahl}
  \emph {et~al.}}]{Stahl:2012exa}%
  \BibitemOpen
  \bibfield  {author} {\bibinfo {author} {\bibfnamefont {A.}~\bibnamefont
  {Stahl}} \emph {et~al.},\ }\href@noop {} {\enquote {\bibinfo {title}
  {Expression of interest for a very long baseline neutrino oscillation
  experiment ({LBNO})},}\ }\bibinfo {howpublished} {CERN-PH-EP-2012-021,
  SPSC-EOI-007} (\bibinfo {year} {2012})\BibitemShut {NoStop}%
\bibitem [{\citenamefont {Abe}\ \emph {et~al.}(2014{\natexlab{b}})\citenamefont
  {Abe} \emph {et~al.}}]{T2Kbeamnue1}%
  \BibitemOpen
  \bibfield  {author} {\bibinfo {author} {\bibfnamefont {K.}~\bibnamefont
  {Abe}} \emph {et~al.} (\bibinfo {collaboration} {T2K collaboration}),\
  }\href@noop {} {\bibfield  {journal} {\bibinfo  {journal} {Phys. Rev. D}\
  }\textbf {\bibinfo {volume} {89}} (\bibinfo {year} {2014}{\natexlab{b}})},\
  \Eprint {http://arxiv.org/abs/1403.2552} {arXiv:1403.2552 [hep-ex]}
  \BibitemShut {NoStop}%
\bibitem [{\citenamefont {Assylbekov}\ \emph {et~al.}(2012)\citenamefont
  {Assylbekov} \emph {et~al.}}]{Assylbekov201248}%
  \BibitemOpen
  \bibfield  {author} {\bibinfo {author} {\bibfnamefont {S.}~\bibnamefont
  {Assylbekov}} \emph {et~al.} (\bibinfo {collaboration} {T2K ND280 P0D
  Collaboration}),\ }\href {\doibase 10.1016/j.nima.2012.05.028} {\bibfield
  {journal} {\bibinfo  {journal} {Nucl.Instrum.Meth.}\ }\textbf {\bibinfo
  {volume} {A686}},\ \bibinfo {pages} {48 } (\bibinfo {year}
  {2012})}\BibitemShut {NoStop}%
\bibitem [{\citenamefont {Tice}\ \emph {et~al.}(2014)\citenamefont {Tice} \emph
  {et~al.}}]{BGTice:2014}%
  \BibitemOpen
  \bibfield  {author} {\bibinfo {author} {\bibfnamefont {B.~G.}\ \bibnamefont
  {Tice}} \emph {et~al.} (\bibinfo {collaboration} {Minerva}),\ }\href@noop {}
  {\bibfield  {journal} {\bibinfo  {journal} {Phys. Rev. Lett}\ }\textbf
  {\bibinfo {volume} {112}},\ \bibinfo {pages} {23181} (\bibinfo {year}
  {2014})},\ \Eprint {http://arxiv.org/abs/1403.2103} {arXiv:1403.2103
  [hep-ex]} \BibitemShut {NoStop}%
\bibitem [{\citenamefont {Tice}(2014)}]{BGTiceThesis}%
  \BibitemOpen
  \bibfield  {author} {\bibinfo {author} {\bibfnamefont {B.}~\bibnamefont
  {Tice}},\ }\emph {\bibinfo {title} {Measurement of Nuclear Dependence in
  Inclusive Charged Current Neutrino Scattering}},\ \href@noop {} {Ph.D.
  thesis},\ \bibinfo  {school} {Rutgers University} (\bibinfo {year}
  {2014})\BibitemShut {NoStop}%
\bibitem [{\citenamefont {Day}\ and\ \citenamefont
  {McFarland}(2012)}]{Day:2012gb}%
  \BibitemOpen
  \bibfield  {author} {\bibinfo {author} {\bibfnamefont {M.}~\bibnamefont
  {Day}}\ and\ \bibinfo {author} {\bibfnamefont {K.~S.}\ \bibnamefont
  {McFarland}},\ }\href {\doibase 10.1103/PhysRevD.86.053003} {\bibfield
  {journal} {\bibinfo  {journal} {Phys. Rev. D}\ }\textbf {\bibinfo {volume}
  {86}},\ \bibinfo {pages} {053003} (\bibinfo {year} {2012})}\BibitemShut
  {NoStop}%
\bibitem [{\citenamefont {Zeller}(2014)}]{Zeller:2014}%
  \BibitemOpen
  \bibfield  {author} {\bibinfo {author} {\bibfnamefont {G.}~\bibnamefont
  {Zeller}},\ }\href@noop {} {\bibfield  {journal} {\bibinfo  {journal}
  {Chinese Physics C}\ }\textbf {\bibinfo {volume} {38}},\ \bibinfo {pages}
  {526} (\bibinfo {year} {2014})}\BibitemShut {NoStop}%
\bibitem [{\citenamefont {Gran}\ \emph {et~al.}(2006)\citenamefont {Gran} \emph
  {et~al.}}]{K2KQE}%
  \BibitemOpen
  \bibfield  {author} {\bibinfo {author} {\bibfnamefont {G.}~\bibnamefont
  {Gran}} \emph {et~al.},\ }\href@noop {} {\bibfield  {journal} {\bibinfo
  {journal} {Phys. Rev. D}\ }\textbf {\bibinfo {volume} {74}},\ \bibinfo
  {pages} {052002} (\bibinfo {year} {2006})}\BibitemShut {NoStop}%
\bibitem [{\citenamefont {Mariani}\ \emph {et~al.}(2011)\citenamefont {Mariani}
  \emph {et~al.}}]{K2KPi1}%
  \BibitemOpen
  \bibfield  {author} {\bibinfo {author} {\bibfnamefont {C.}~\bibnamefont
  {Mariani}} \emph {et~al.},\ }\href@noop {} {\bibfield  {journal} {\bibinfo
  {journal} {Phys. Rev. D}\ }\textbf {\bibinfo {volume} {83}},\ \bibinfo
  {pages} {054023} (\bibinfo {year} {2011})}\BibitemShut {NoStop}%
\bibitem [{\citenamefont {Rodriguez}\ \emph {et~al.}(2008)\citenamefont
  {Rodriguez} \emph {et~al.}}]{K2KPi2}%
  \BibitemOpen
  \bibfield  {author} {\bibinfo {author} {\bibfnamefont {A.}~\bibnamefont
  {Rodriguez}} \emph {et~al.},\ }\href@noop {} {\bibfield  {journal} {\bibinfo
  {journal} {Phys. Rev. D}\ }\textbf {\bibinfo {volume} {78}},\ \bibinfo
  {pages} {032003} (\bibinfo {year} {2008})}\BibitemShut {NoStop}%
\bibitem [{\citenamefont {Nakayama}\ \emph {et~al.}(2005)\citenamefont
  {Nakayama} \emph {et~al.}}]{K2KPi3}%
  \BibitemOpen
  \bibfield  {author} {\bibinfo {author} {\bibfnamefont {S.}~\bibnamefont
  {Nakayama}} \emph {et~al.},\ }\href@noop {} {\bibfield  {journal} {\bibinfo
  {journal} {Phys. Rev. Lett.}\ }\textbf {\bibinfo {volume} {619}},\ \bibinfo
  {pages} {255} (\bibinfo {year} {2005})}\BibitemShut {NoStop}%
\bibitem [{\citenamefont {Hasegawa}\ \emph {et~al.}(2005)\citenamefont
  {Hasegawa} \emph {et~al.}}]{K2KPi4}%
  \BibitemOpen
  \bibfield  {author} {\bibinfo {author} {\bibfnamefont {M.}~\bibnamefont
  {Hasegawa}} \emph {et~al.},\ }\href@noop {} {\bibfield  {journal} {\bibinfo
  {journal} {Phys. Rev. Lett.}\ }\textbf {\bibinfo {volume} {95}},\ \bibinfo
  {pages} {252301} (\bibinfo {year} {2005})}\BibitemShut {NoStop}%
\bibitem [{\citenamefont {Abe}\ \emph {et~al.}(2014{\natexlab{c}})\citenamefont
  {Abe} \emph {et~al.}}]{PhysRevD.89.092003}%
  \BibitemOpen
  \bibfield  {author} {\bibinfo {author} {\bibfnamefont {K.}~\bibnamefont
  {Abe}} \emph {et~al.} (\bibinfo {collaboration} {The T2K Collaboration}),\
  }\href {\doibase 10.1103/PhysRevD.89.092003} {\bibfield  {journal} {\bibinfo
  {journal} {Phys. Rev. D}\ }\textbf {\bibinfo {volume} {89}},\ \bibinfo
  {pages} {092003} (\bibinfo {year} {2014}{\natexlab{c}})}\BibitemShut
  {NoStop}%
\bibitem [{\citenamefont {Schalicke}\ \emph {et~al.}(2011)\citenamefont
  {Schalicke} \emph {et~al.}}]{GEANT4_1}%
  \BibitemOpen
  \bibfield  {author} {\bibinfo {author} {\bibfnamefont {A.}~\bibnamefont
  {Schalicke}} \emph {et~al.},\ }\href@noop {} {\bibfield  {journal} {\bibinfo
  {journal} {J. Phys: Conf. Ser.}\ }\textbf {\bibinfo {volume} {331}} (\bibinfo
  {year} {2011})}\BibitemShut {NoStop}%
\bibitem [{\citenamefont {Kadri}\ \emph {et~al.}(2007)\citenamefont {Kadri}
  \emph {et~al.}}]{GEANT4_2}%
  \BibitemOpen
  \bibfield  {author} {\bibinfo {author} {\bibfnamefont {O.}~\bibnamefont
  {Kadri}} \emph {et~al.},\ }\href@noop {} {\bibfield  {journal} {\bibinfo
  {journal} {Nucl. Instrum. Meth}\ }\textbf {\bibinfo {volume} {B258}},\
  \bibinfo {pages} {381} (\bibinfo {year} {2007})}\BibitemShut {NoStop}%
\bibitem [{\citenamefont {Sawkey}\ and\ \citenamefont
  {Faddegon}(2009)}]{GEANT4_3}%
  \BibitemOpen
  \bibfield  {author} {\bibinfo {author} {\bibfnamefont {D.~L.}\ \bibnamefont
  {Sawkey}}\ and\ \bibinfo {author} {\bibfnamefont {B.~A.}\ \bibnamefont
  {Faddegon}},\ }\href@noop {} {\bibfield  {journal} {\bibinfo  {journal} {Med
  Phys.}\ }\textbf {\bibinfo {volume} {36(3)}},\ \bibinfo {pages} {698}
  (\bibinfo {year} {2009})}\BibitemShut {NoStop}%
\bibitem [{\citenamefont {Ivanchenko}(2003)}]{GEANT4_4}%
  \BibitemOpen
  \bibfield  {author} {\bibinfo {author} {\bibfnamefont {V.~N.}\ \bibnamefont
  {Ivanchenko}},\ }\href@noop {} {\bibfield  {journal} {\bibinfo  {journal}
  {Nucl. Instrum. Meth}\ }\textbf {\bibinfo {volume} {A502}},\ \bibinfo {pages}
  {666} (\bibinfo {year} {2003})}\BibitemShut {NoStop}%
\bibitem [{\citenamefont {Kolmogorov}(1933)}]{Kolmogorov}%
  \BibitemOpen
  \bibfield  {author} {\bibinfo {author} {\bibfnamefont {A.}~\bibnamefont
  {Kolmogorov}},\ }\href@noop {} {\bibfield  {journal} {\bibinfo  {journal} {G.
  Ist. Ital. Attuari}\ }\textbf {\bibinfo {volume} {4}},\ \bibinfo {pages}
  {83–91} (\bibinfo {year} {1933})}\BibitemShut {NoStop}%
\bibitem [{\citenamefont {Smirnov}(1948)}]{Smirnov}%
  \BibitemOpen
  \bibfield  {author} {\bibinfo {author} {\bibfnamefont {N.}~\bibnamefont
  {Smirnov}},\ }\href {\doibase doi:10.1214/aoms/1177730256} {\bibfield
  {journal} {\bibinfo  {journal} {Annals of Mathematical Statistics}\ }\textbf
  {\bibinfo {volume} {19}},\ \bibinfo {pages} {279–281} (\bibinfo {year}
  {1948})}\BibitemShut {NoStop}%
\end{thebibliography}%

\end{document}